\documentclass[singlecolumn,superscriptaddress]{revtex4}
\usepackage{amsmath}
\usepackage{amssymb}
\usepackage{amsfonts}
\usepackage{graphicx}
\usepackage{bbold}
\usepackage{bm}
\usepackage{bm}
\usepackage{cancel}
\usepackage{times,float}
\usepackage{graphicx}
\usepackage[usenames,dvipsnames,svgnames]{xcolor}
\usepackage{hyperref}
\usepackage{comment}
\hypersetup{colorlinks=true, linkcolor=Blue, citecolor=PineGreen,urlcolor=Blue}
\usepackage{multirow}
\usepackage{ulem}
\usepackage{color,epsfig}
\usepackage{bm}
\usepackage{slashed}
\usepackage{hyperref}
\usepackage{subfigure}
\usepackage{feynmp}
\usepackage{hyperref}
\usepackage{etoolbox}
\usepackage{orcidlink}
\hypersetup{
    colorlinks=true,
    linkcolor=Blue,
    citecolor=Blue,
    urlcolor=Blue
}

\newcommand{\bea}{\begin{eqnarray}}
\newcommand{\eea}{\end{eqnarray}}
\newcommand{\be}{\begin{equation}}
\newcommand{\ee}{\end{equation}}

\newcommand{\sgn}{{\rm sign}}

\renewcommand\vec{\bm}

\newcommand{\nn}{\nonumber}
\newcommand{\ii}{\mathrm{i}}

\newcommand{\qB}{eB}

\newcommand{\sign}{\text{sign}(\qB)}

\DeclareMathOperator*{\SumInt}{%
\mathchoice%
{\ooalign{$\displaystyle\sum$\cr\hidewidth$\displaystyle\int$\hidewidth\cr}}
  {\ooalign{\raisebox{.14\height}{\scalebox{.7}{$\textstyle\sum$}}\cr\hidewidth$\textstyle\int$\hidewidth\cr}}
  {\ooalign{\raisebox{.2\height}{\scalebox{.6}{$\scriptstyle\sum$}}\cr$\scriptstyle\int$\cr}}
  {\ooalign{\raisebox{.2\height}{\scalebox{.6}{$\scriptstyle\sum$}}\cr$\scriptstyle\int$\cr}}
}

\begin{document}

\title{QED vertex and anomalous magnetic moment in the presence of a magnetic field}
\author{Alejandro Ayala~\orcidlink{0000-0003-3929-9209}}
\email{ayala@nucleares.unam.mx}
\affiliation{Instituto de Ciencias
Nucleares, Universidad Nacional Aut\'onoma de M\'exico, Apartado
Postal 70-543, CdMx 04510,
Mexico}
\author{Enrique Mu\~noz~\orcidlink{0000-0003-4457-0817}}
\email
[Corresponding author: ]{ejmunozt@uc.cl}
\affiliation{Facultad de F\'isica, Pontificia Universidad Cat\'olica de Chile, Vicu\~{n}a Mackenna 4860, Santiago, Chile}
\author{Marcelo Loewe~\orcidlink{0000-0002-8998-6923}}
\email{marcelo.loewe@uss.cl}
\affiliation{Facultad de Ingenier\'\i a, Universidad San Sebasti\'an, Bellavista 7, Recoleta, Santiago, Chile}
\affiliation{Centre for Theoretical and Mathematical Physics, and Department of Physics, University of Cape Town, Rondebosch 7700, South Africa}
\author{Juan Cristóbal Rojas
\orcidlink{0000-0002-3888-7538}}
\email{jurojas@ucn.cl}
\affiliation{Departamento de Física, Universidad Católica del Norte, Angamos 0610, Antofagasta, Chile}
\author{Norberto Scoccola}
\email{scoccola@gmail.com}
\affiliation{Departamento de F\'isica Te\'orica,
Comisi\'on Nacional de Energ\'ia At\'omica, Av. Libertador 8250, (1429) Buenos Aires, Argentina}
\affiliation{CONICET, Rivadavia 1917, (1033) Buenos Aires, Argentina}


\begin{abstract}

We compute the fermion-photon vertex in QED in the presence of a constant and uniform magnetic background up to one-loop order. We show that even at tree-level, the vertex is modified due to the loss of Lorentz invariance induced by the magnetic field, thus breaking into longitudinal and transverse pieces. Moreover, the radiative corrections induce the emergence of a rich tensor structure that includes the anomalous magnetic moments in the transverse, parallel, and mixed transverse/parallel directions. We concentrate on studying one of these anomalous magnetic moment components, the one in the purely transverse direction. We find the selection rules for transitions between a few low-lying Landau levels and show that the amplitudes for transitions from an initial to a final Landau level differ by a sign from the reverse process due to the loss of time reversal invariance induced by the presence of the field. Contrary to the vacuum case, the amplitudes are, in general, complex, and the phase factor can be interpreted in terms of a finite life-time of the decaying state. For the anomalous magnetic moment in the purely transverse direction, transitions between states occupying both the lowest Landau levels are forbidden. Moreover, for the computation of the allowed transitions, we find that it is not necessary to include a photon mass since the magnetic field acts as an infrared regulator.

\end{abstract}

\maketitle 

\section{Introduction}

In quantum field theories, loop corrections are an essential ingredient for exploring the properties of interactions as functions of either the momentum exchanged in a given process~\cite{CMS:2024trs} or possible changes in environmental variables~\cite{Steffens:2004sg,Ayala:2025qag}. In Quantum Electrodynamics (QED), Schwinger's pioneering one-loop calculation~\cite{Schwinger:1948iu} revealed the existence of the electron's anomalous magnetic moment, a prediction that has since been confirmed with extraordinary precision by experiment~\cite{Fan:2022eto}. The anomalous magnetic moment thus stands as one of the most remarkable manifestations of quantum fluctuations and provides a natural framework for investigating how radiative corrections are modified when the vacuum is replaced by a nontrivial background.

In recent years, considerable attention has been devoted to understanding how magnetic fields modify particle properties within both effective and fundamental theories. These studies have shown, in particular, that particle masses and interaction strengths become field dependent, evolving with the magnetic field intensity~\cite{Gusynin:1999pq,Mueller:2014tea,Avancini:2015ady,Avancini:2016fgq,Avancini:2018svs,Ayala:2015qwa,Carlomagno:2022arc,Carlomagno:2022inu,Ayala:2018zat,Avancini:2021pmi,Coppola:2018vkw,Coppola:2023mmq,Li:2016dta,Ayala:2006sv,Ayala:2019akk,Ayala:2021lor,Rojas:2008sg,Castano-Yepes:2022luw,Castano-Yepes:2023brq,Castano-Yepes:2024ltr,Moreira:2022dwo,Castano-Yepes:2024ctr,Wang:2026xsm,Adhikari:2024bfa}. The presence of a magnetic background introduces a preferred spatial direction, thereby reducing the full Lorentz symmetry of the system. As a consequence, charge conjugation and time-reversal symmetries are explicitly broken, and space-time naturally decomposes into longitudinal and transverse sectors relative to the field direction. The anisotropic structure is inherited by all Lorentz-covariant quantities, including propagators, self-energies, vertex functions, and other scalar, vector, and tensor operators, which must be expressed in terms of independent components parallel and perpendicular to the magnetic field. One manifestation of this symmetry breaking is the emergence of distinct particle propagation velocities in the longitudinal and transverse directions~\cite{Fayazbakhsh:2012vr,Sheng:2020hge}. Likewise, even at tree-level, the fermion–photon interaction vertex separates into longitudinal and transverse components~\cite{Miransky:2015ava}. Beyond tree-level, radiative corrections generate additional anisotropic structures, including different contributions to the anomalous magnetic moment (AMM). In general, the fermion magnetic moment splits into three independent components. Explicit calculations of these contributions have been carried out both in the lowest-Landau-level (LLL) approximation~\cite{Fraga:2024klm} and in the weak-field regime~\cite{Lin:2021bqv}. In the LLL it was shown that only the fully longitudinal component of the anomalous magnetic moment (AMM) is generated. In the weak field regime, it has been shown that the AMM receives a negative, magnetic field dependent contribution. An even earlier result for the magnetic field-induced modification of the electron AMM was found in Ref.~\cite{Baier:1975uj}. Consequences of a magnetic field-dependent quark AMM have been explored in Refs.~\cite{Tavares:2023oln,Farias:2021fci,Fayazbakhsh:2014mca}.

In the presence of a magnetic background, the motion of charged particles differs significantly from that in vacuum. Consequently, the states describing their space-time propagation are no longer plane waves but are instead given by the Ritus eigenfunctions~\cite{Ritus:1972ky,Ritus:1978cj}, characterized by discrete quantum numbers known as Landau levels. The momentum component that becomes quantized depends on the particular gauge choice. Therefore, calculations involving on-shell charged particles must be carried out using these Ritus eigenfunctions. 

In this work, we compute the magnetic-field-induced modifications to the fermion-photon interaction vertex in QED at both leading (tree-level) and next-to-leading (one-loop) order for on-shell fermions occupying specific Landau levels and for arbitrary magnetic-field strengths. We identify the most general vector and tensor structures that emerge as a consequence of the magnetic background. From this decomposition, we isolate the magnetic-field-induced contribution to the transverse anomalous magnetic moment (AMM), which, to the best of our knowledge, has not been reported previously. We also derive the selection rules governing transitions between external fermion states occupying given Landau levels and present explicit expressions for several low-lying transition amplitudes contributing to the transverse AMM. Furthermore, we show that transitions between fermion Landau levels $k$ and $k'$ generally depend on the ordering of the initial and final states. This asymmetry is a direct manifestation of the breaking of time-reversal symmetry by the magnetic background. The work is organized as follows: In Sec.~\ref{sec2}, we set up the calculation for the fermion-photon vertex in a background magnetic field of arbitrary strength in the Ritus formalism. We discuss the leading order corrections in Sec.~\ref{sec2A} and the one-loop correction in Sec.~\ref{sec2B}. In Sec.~\ref{sec3} we specialize to computing the anomalous magnetic moment in the purely transverse direction and discuss the selection rules for transitions between different low-lying Landau levels. We show that this form factor vanishes for transitions between external fermions occupying the LLL. We also show that, unlike the case of the purely longitudinal anomalous magnetic moment form factor, the purely transverse form factor is infrared safe. We finally conclude in Sec.~\ref{concl} and reserve the details of the calculation for the appendices.

\section{The fermion-photon interaction vertex in a background magnetic field}\label{sec2}

\begin{figure}[t!]
    \centering
    \includegraphics[width=0.8\linewidth]{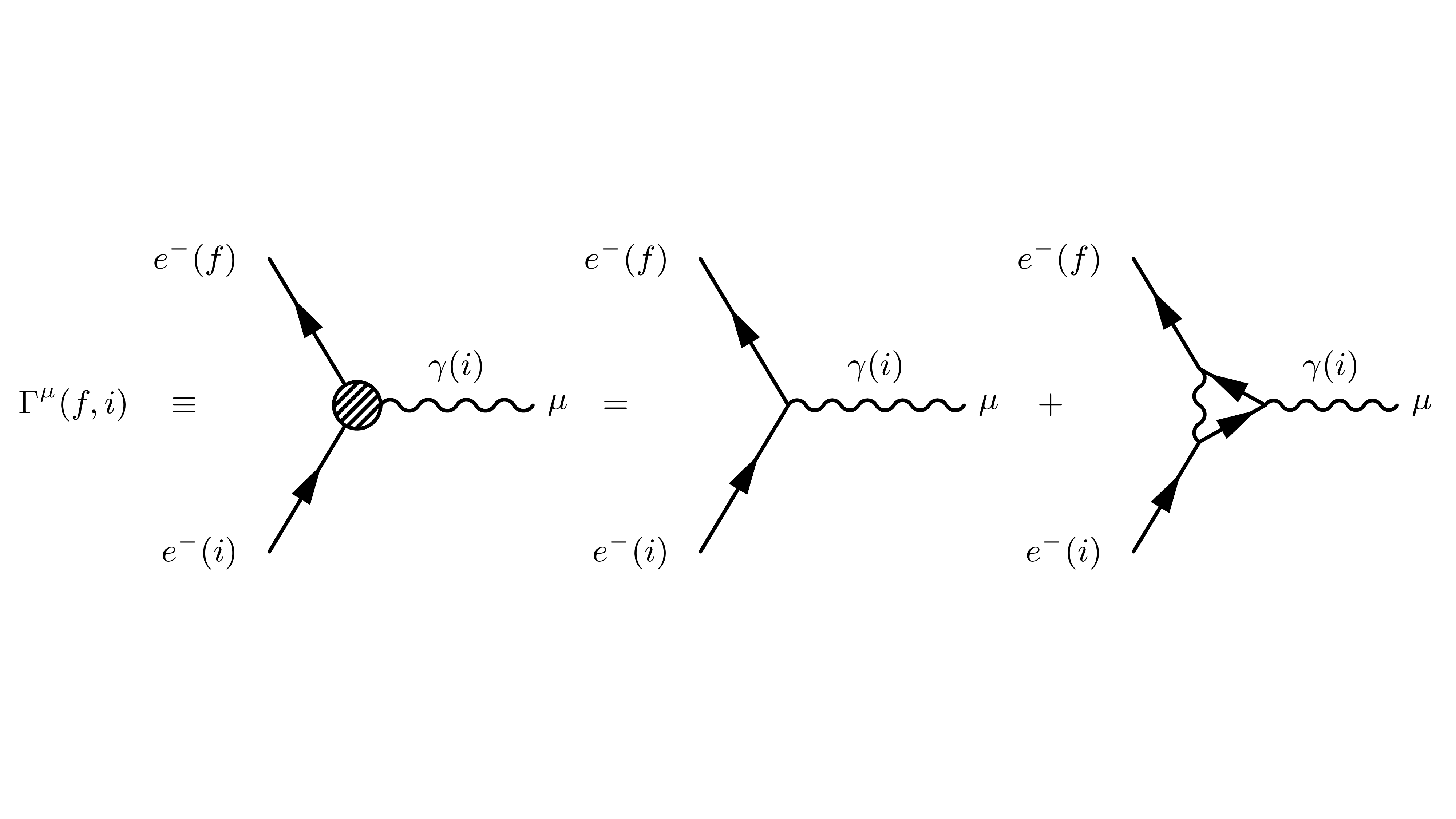}
    
    \caption{Feynman diagrams contributing to the fermion-photon interaction vertex up to one-loop order.}
    \label{fig1}
\end{figure}

Let us start by considering the general definition of the amplitude for the process $e^{-} + \gamma \rightarrow e^{-}$, involving the scattering of an electron with a photon, 
\begin{eqnarray}
\langle e^{-}(f)|e^{i\int d^4 x \mathcal{L}_{\rm{int}}(x)}|e^{-}(i),\gamma(i)\rangle &=& \langle e^{-}(f)|i\int_x  \mathcal{L}_{\rm{int}}(x)|e^{-}(i),\gamma(i)\rangle + \langle e^{-}(f)|\frac{i^3}{3!}\left[\int_x \mathcal{L}_{\rm{int}}(x)\right]^3|e^{-}(i),\gamma(i)\rangle + \ldots\nonumber\\
&=& \mathcal{V}_{LO}(f,i) + \mathcal{V}_{NLO}(f,i) + \ldots
\end{eqnarray}
where $|e^{-}(i)\gamma(i)\rangle$ and $|e^{-}(f)\rangle$ represent the in and out states, respectively, while 
\begin{eqnarray}
\mathcal{L}_{int}(x) = -e \overline{\psi}(x)\gamma^{\mu}A_{\mu}(x)\psi(x),
\end{eqnarray}
is the fermion-photon interaction Lagrangian density. Here, we defined
$\mathcal{V}_{LO}(i,f)$ and $\mathcal{V}_{NLO}(i,f)$ as the leading order (LO) and next to leading order (NLO) contributions to the scattering amplitude, respectively. These two contributions are represented by the Feynman diagrams depicted in Fig.~\ref{fig1}.

Since we are interested in the process under the presence of a classical, constant background magnetic field $\mathbf{B} = \hat{z} B = \nabla\times\mathbf{A}_{BG}$ along the $\hat{z}$-direction, we shall describe the fermion operators, and their subsequent propagators, in the Ritus eigenbasis, by choosing the Landau Gauge 2 (LG2): $A_{BG}^{\mu}(x) = (0,0,B x^{1},0)$. For this purpose, we introduce some basic definitions related to the Ritus basis. For a particle with electric charge $e$, we have the magnetic field-dependent variables
\be
B_e = |e B|,\,\,\,
s = \sgn(eB),
\label{eq:eB}
\ee
and the energy spectrum $E_e = \sqrt{m^2 + 2 k B_e + \left( q^3 \right)^2}$.

For notational purposes, in the LG2 we also introduce the 3-component vector $\breve{q} = (k,q^2,q^3)$,
with $k \in\mathbb{N}_0$ the Landau level index, and $q^i\in\mathbb{R}$ ($i=2,3$), the continuum 3-momentum components.
For the linear superposition over Ritus eigenfunctions, we use the symbol (in LG2)
\bea
\SumInt_{\{ \overline{q}_{E_e} \}} \equiv \sum_{k=0}^{\infty}\int\frac{dq^0 dq^2 dq^3}{(2\pi)^3} (2\pi )\delta\left(  q^0 - E_e\right).
\eea
The electron spinor field in the Ritus basis is then given by the superposition
\bea
\psi(x) = \SumInt_{\{ \overline{q}_{E_e} \}} \sum_{a=1,2}\frac{1}{2 E_e}\Bigg\{ b(\breve{q},a) U(x,\overline{q},a)+ d^{\dagger}(\breve{q},a) V(x,\overline{q},a)
\Bigg\}.
\label{eq:spinor}
\eea
Here, the spinor amplitudes for polarization $a=1,2$ are defined by
\bea
U(x,\overline{q},a) &=& \mathbb{E}^e(x,\overline{q}) \left.u_e (q_{\parallel},k,a)\right|_{q^0=E_e}\nn\\
V(x,\overline{q},a) &=& \tilde{\mathbb{E}}^{-e}(x,\overline{q}) \left.v_{-e} (q_{\parallel},k,a)\right|_{q^0=E_e}\ ,
\label{eq_URit}
\eea
where, in terms of the spin projectors $\Delta^{\lambda} = \frac{1}{2}\left( 1 + \ii \lambda  \gamma^1 \gamma^2\right)$ (for $\lambda=\pm$), we define the Ritus functions
\bea
\mathbb{E}^e(x,\overline{q}) &=& \sum_{\lambda=\pm}\Delta^{\lambda}\mathcal{F}_e (x,\overline{q}_{\lambda})\nn\\
\tilde{\mathbb{E}}^{-e}(x,\overline{q}) &=& \sum_{\lambda=\pm}\Delta^{\lambda}\mathcal{F}_{-e}^{*} (x,\overline{q}_{-\lambda}),
\label{eq_ERitus}
\eea
whose arguments are the four component vectors $\overline{q}_{\lambda} \equiv \left( q^0, k_{s\lambda}, q^2, q^3 \right)$, for the spin-shifted Landau levels ($s = \sign$, $\lambda=\pm$)
\bea
k_{s\lambda} \equiv k - \frac{(1 - s\lambda)}{2}.
\eea
Finally, in the LG2 the functions in Eq.~\eqref{eq_ERitus} are defined by
\bea
\mathcal{F}_{Q}(x,\overline{q}) = N_k e^{-\ii\left( q^0 x^0 - q^2 x^2 - q^3 x^3 \right)} D_k \left( \sqrt{2 B_Q}\left( x^1 - s\frac{q^2}{B_Q} \right) \right),
\label{eq_FQ}
\eea
where $D_k(\xi)$ are parabolic cylinder functions $D_k(\xi) = 2^{-k} e^{-\frac{\xi^2}{4}} H_k (\xi/\sqrt{2})$, with normalization coefficients $N_k = \frac{1}{\sqrt{k!}} \left( 4\pi B_Q \right)^{1/4}$. As usual, we use the convention $H_{-1}(z) = 0$.
We shall also define the spinors, in the Ritus eigenbasis, which are solutions of the Dirac equation as follows
\bea
\left( \slashed{\Pi}_s(p_{\parallel},n) - m \right) \left.u_e(p_{\parallel},n,a)\right|_{p^0 = E_e} &=& 0\nn\\
\left( \slashed{\Pi}_{-s}(p_{\parallel},n) + m \right) \left.v_{-e}(p_{\parallel},n,a)\right|_{p^0=E_e} &=& 0.
\label{eq_Dirac_Ritus1}
\eea
Here, we defined the \lq\lq canonical momenta"
\bea
\Pi^{\mu}_s(p_{\parallel},n) = \left( p^0,0,-s\sqrt{2 n B_e},p^3  \right), 
\eea
such that we can decompose the matrix structure as follows
\bea
\slashed{\Pi}_s(p_{\parallel},n) &\equiv& \slashed{\Pi}_{\parallel}^{s}(p_{\parallel}) + \slashed{\Pi}_{\perp}^{s}(n)\nn\\
&=& \slashed{p}_{\parallel} -s\sqrt{2 B_e n}\gamma^{2}.
\label{eq_Pisep}
\eea
The spinors are explicitly defined by
\bea
\left.u_e(p_{\parallel},n,a)\right|_{p^0=E_e} &=& \frac{1}{\sqrt{2(E_e + m)}}\left[ \left.\slashed{\Pi}_s(p_{\parallel},n)\right|_{p^0=E_e} + m\right]\left( \begin{array}{c}\phi^{(a)}\\\phi^{(a)} \end{array}\right)\nn\\
\left.v_{-e}(p_{\parallel},n,a)\right|_{p^0=E_e} &=& \frac{1}{\sqrt{2(E_e + m)}}\left[ \left.-\slashed{\Pi}_{-s}(p_{\parallel},n)\right|_{p^0=E_e} + m\right]\left( \begin{array}{c}\tilde{\phi}^{(a)}\\-\tilde{\phi}^{(a)} \end{array}\right).
\eea
The bispinor components above are $\phi^{(1)\dagger} = -\tilde{\phi}^{(2)\dagger} = (1,0)$, and $\phi^{(2)\dagger} = \tilde{\phi}^{(1)\dagger} = (0,1)$.

A property that we shall often use is related to an alternative way to express the Dirac equation, as satisfied by the spinor eigenfunctions defined in Eq.~\eqref{eq_Dirac_Ritus1}. By combining Eq.~\eqref{eq_Pisep} with Eq.~\eqref{eq_Dirac_Ritus1}, we have
\bea
\left( \slashed{\Pi}_s(p_{\parallel},n) - m \right)\left.u_e(p_{\parallel},n,a)\right|_{p^0=E_e} = \left( \slashed{p}_{\parallel} - s\sqrt{2 B_e n}\gamma^2 - m\right)\left.u_e(p_{\parallel},n,a)\right|_{p^0=E_e} = 0,
\eea
thus proving the identity
\bea
\slashed{p}_{\parallel}\left.u_e(p_{\parallel},n,a)\right|_{p^0=E_e} = \left( m + s\sqrt{2 B_e n}\gamma^2 \right) \left.u_e(p_{\parallel},n,a)\right|_{p^0=E_e}.
\label{eq_Dirac_Ritus2}
\eea
Similarly, we also obtain
\bea
\left.\overline{u}_e (p_{\parallel},n,a)\right|_{p^0=E_e} \slashed{p}_{\parallel} = \left.\overline{u}_e (p_{\parallel},n,a)\right|_{p^0=E_e}\left( m + s\sqrt{2 B_e n}\gamma^2 \right).
\label{eq_Dirac_Ritus3}
\eea
Finally, we remark that the Schwinger propagator admits a representation in terms of the Ritus eigenbasis as follows
\bea
S_F(x,y) = \sum_{n}\int \frac{d^3 \tilde{q}}{(2\pi)^3} \mathbb{E}^e(x,\tilde{q},n)\frac{\slashed{\Pi}_s(q_{\parallel},n) + m}{q_{\parallel}^2 - 2 n B_e - m^2}\tilde{\mathbb{E}}^{e}(y,\tilde{q},n),
\label{eq_Ritusprop}
\eea
where $\tilde{q} = (q^0,q^2,q^3)$.
\subsection{The Vertex at the tree-level (leading order)}\label{sec2A}

\begin{figure}[t!]
    \centering
    \includegraphics[width=0.5\linewidth]{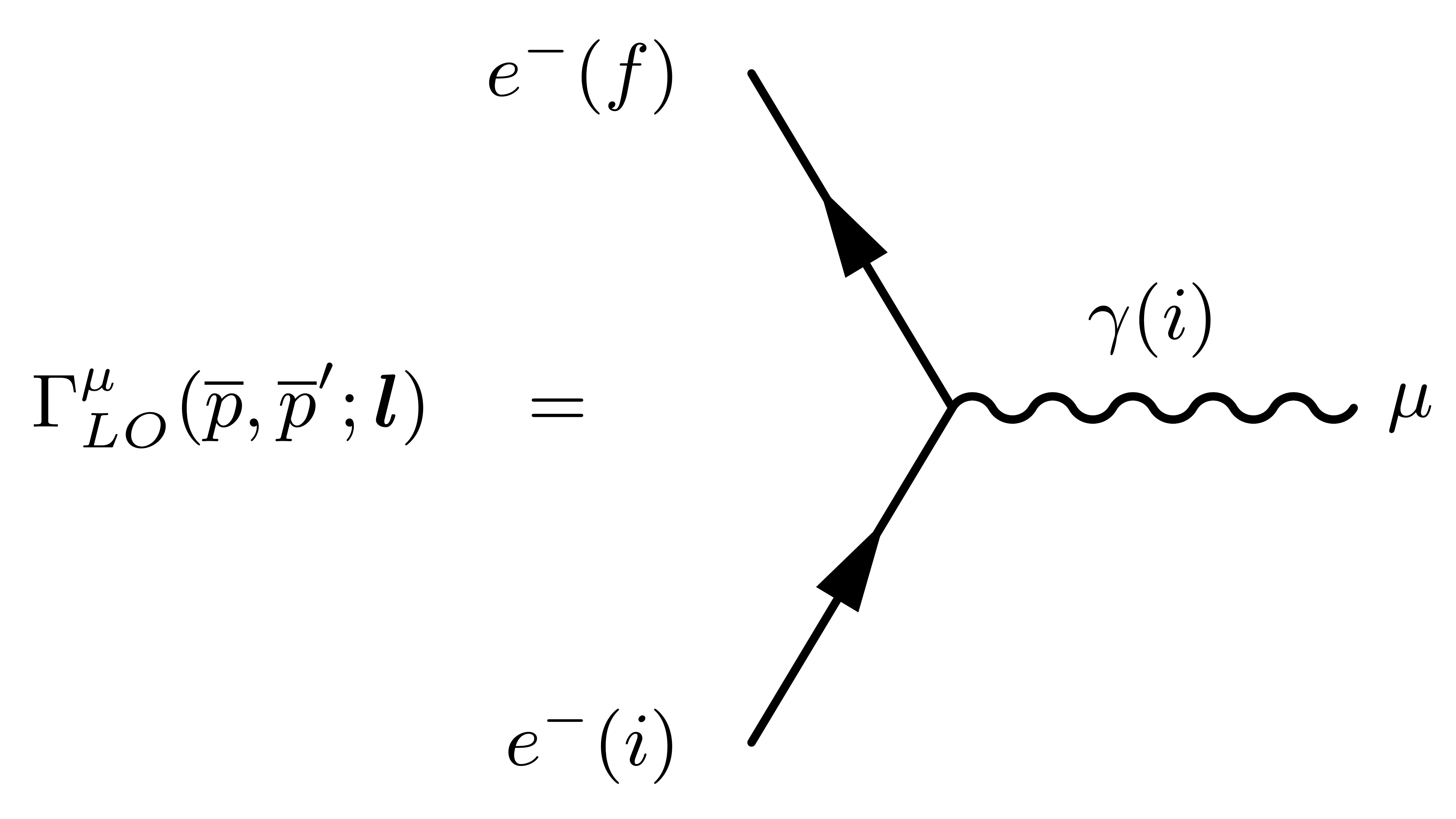}
    
    \caption{Feynman diagram for the tree-level (LO) contribution to the fermion-photon vertex.}
    \label{fig2}
\end{figure}

We proceed to the leading order (LO) calculation of the vertex. Details can be found in Appendix~\ref{LO}. At LO, the scattering process $e^{-} + \gamma\rightarrow e^{-}$ 
is represented by the Feynman diagram depicted in Fig.~\ref{fig2},
whose corresponding amplitude is given by the expression
\bea
\mathcal{V}_{LO}(\breve{p},a,\breve{p}',a';\vec{l},c) &=& \langle e^{-}(\breve{p},a)|\ii \int_{x}\mathcal{L}_{int}(x)|\gamma(\vec{l},c)e^{-}(\breve{p}',a')\rangle\nn\\
&=& -\ii e \int d^4 x \langle 0|b(\breve{p},a) \overline{\psi}(x)\gamma^{\mu}\psi(x)A_{\mu}(x)b^{\dagger}(\breve{p}',a')a^{\dagger}(\vec{l},c)|0\rangle.
\label{eq:VLO0}
\eea
At this point, we need to consider the expansion of the spinors according to Eq.~\eqref{eq:spinor}, 
as well as the representation of the photon gauge field
\bea
A^{\mu}(x) = \int_{l'_{E_{\gamma}}}\frac{1}{2 E'_{\gamma}}\sum_{c'}\left\{
a(\vec{l}',c')\epsilon^{\mu}(\vec{l'},c')e^{-\ii l'\cdot x}+a^{\dagger}(\vec{l},c)\left[ \epsilon^{\mu}(\vec{l'},c')\right]^* e^{\ii l'\cdot x}
\right\}.
\eea
Inserting these into Eq.~\eqref{eq:VLO0}, we recognize that the nonzero terms are the following
\bea
\mathcal{V}_{LO}(\breve{p},a,\breve{p}',a',\vec{l},c) &\!\!\!=\!\!\!& -\ii e \int d^4 x \SumInt_{\{ \overline{q}_{E_{e,1}} \}} \sum_{a_1=1,2}\frac{1}{2 E_{e,1}} \SumInt_{\{ \overline{q}_{E_{e,2}} \}} \sum_{a_2=1,2}\frac{1}{2 E_{e,2}} \int_{l'_{E_{\gamma}}}\frac{1}{2 E'_{\gamma}}\sum_{c'}\epsilon_{\mu}(\vec{l'},c') \overline{U}(x,\overline{q}_1,a_1)\gamma^{\mu}U(x,\overline{q}_2,a_2)\nn\\
&\!\!\!\times\!\!\!&  e^{-\ii l'\cdot x}\langle 0| b(\breve{p},a)b^{\dagger}(\breve{q}_1,a_1) b(\breve{q}_2,a_2)b^{\dagger}(\breve{p}',a')   |0\rangle
\langle 0| a(\vec{l}',c')a^{\dagger}(\vec{l},c) |0\rangle.
\eea
Here, we calculate the matrix elements applying Wick's theorem, to obtain
\bea
\langle 0| a(\vec{l}',c')a^{\dagger}(\vec{l},c) |0\rangle &=& - g^{cc'}2 E_{\gamma}(2\pi)^3\delta^{(3)}(\vec{l}'-\vec{l})\nn\\
\langle 0| b(\breve{p},a)b^{\dagger}(\breve{q}_1,a_1) b(\breve{q}_2,a_2)b^{\dagger}(\breve{p}',a')   |0\rangle
&=& 2 E_{e,1}\delta_{a_1,a}(2\pi)^{2}\delta_{k_1,k}\delta(q_1^2 - p^2)\delta(q_1^3 - p^3)\nn\\
&&\times 2 E_{e,2}\delta_{a_2,a'}(2\pi)^{2}\delta_{k_2,k'}\delta(q_2^2 - p'^2)\delta(q_2^3 - p'^3).
\label{eq:bbbbaa1}
\eea
Integrating out the delta functions, we end up with the expression
\bea
\mathcal{V}_{LO}(\breve{p},a,\breve{p}',a',\vec{l},c) = -\ii e\, \epsilon_{\mu}(\vec{l},c)\int d^4 x \overline{U}(x,\overline{p},a)\gamma^{\mu}U(x,\overline{p}',a') e^{-\ii l\cdot x}.
\eea
Now, we insert the definition of the spinor amplitudes in the Ritus basis according to Eq.~\eqref{eq_URit},
such that the amplitude reduces to the form
\bea
\mathcal{V}_{LO}(\breve{p},a,\breve{p}',a',\vec{l},c) = -\ii e\, \epsilon_{\mu}(\vec{l},c) \overline{u}_e (k,p^3,a) \Gamma^{\mu}_{LO}(\overline{p},\overline{p}',l)u_e (k',p'^3,a'),
\eea
where we identify the expression for the vertex at the tree-level
\bea
\Gamma^{\mu}_{LO}(\overline{p},\overline{p}',l) = \sum_{\lambda,\lambda'}\Delta^{\lambda}\gamma^{\mu}\Delta^{\lambda'}\int d^4x \mathcal{F}_e^{*}(x,\overline{p}_{\lambda})\mathcal{F}_e(x,\overline{p}'_{\lambda'})e^{-\ii l\cdot x}.
\eea
Now, we apply the following identity relating the Dirac matrices with the spin projectors
\bea
\Delta^{\lambda}\gamma^{\mu}\Delta^{\lambda'} = \Delta^{\lambda}\left[ \delta_{\lambda',\lambda}\gamma_{\parallel}^{\mu}
+ \delta_{\lambda',-\lambda}\gamma_{\perp}^{\mu}\right],
\eea
to obtain our final result for the vertex at tree-level (LO)
\bea
\Gamma^{\mu}_{LO}(\overline{p},\overline{p}',l) &=& (2\pi)^3 \delta^{(3)}( \tilde{l} - \tilde{p} + \tilde{p}' ) \sum_{\lambda = \pm}\Delta^{\lambda}\left\{\gamma^{\mu}_{\parallel} \mathcal{J}_{k_{\lambda},k'_{\lambda}}^{- -}(p^2,p'^2,l^1) + \gamma_{\perp}^{\mu} \mathcal{J}_{k_{\lambda},k'_{-\lambda}}^{- -}(p^2,p'^2,l^1) \right\}.
\label{eq_GLOmu}
\eea
As shown in Appendix~\ref{IntegralsDD}, the above functions are defined by
\bea
\mathcal{J}_{k_{\lambda},k'_{\lambda'}}^{s s}(p^2,p'^2,l^1) &=& e^{\ii \frac{(s p^2 + s p'^2)l^1}{\sqrt{2 B_e}}} \mathcal{G}_{k_{\lambda},k'_{\lambda'}}^{s s}(l^1,l^2 = s p'^2 - s p^2)\nn\\
&=& 2\pi e^{\ii \frac{(s p^2 + s p'^2)l^1}{\sqrt{2 B_e}}} e^{-\frac{l_{\perp}^2}{4 B_e}} \sqrt{\frac{k_{s\lambda}!}{k'_{s\lambda'}!}}\ii^{k'_{s\lambda'}-k_{s\lambda}}e^{\ii(k'_{s\lambda'} - k_{s\lambda})\phi_{\perp}}\left( \frac{l_{\perp}^2}{2 B_e} \right)^{\frac{k'_{s\lambda'}-k_{s\lambda}}{2}}  L_{k_{s\lambda}}^{k'_{s\lambda'}-k_{s\lambda}}\left( \frac{l_{\perp}^2}{2 B_e} \right),
\label{eq_Jfunc}
\eea
where $\tan\phi_{\perp} = \frac{s p'^2 - s p^2}{l^1}$, and $l_{\perp} = (l^1,l^2)$.

A remarkable feature of the vector structure at tree-level for the vertex obtained in Eq.~\eqref{eq_GLOmu} is that the spin polarization of the scattered electron state is preserved $\lambda' = \lambda$ for the parallel vector component $\gamma^{\mu}_{\parallel}$, while it is flipped $\lambda'=-\lambda$ for the transverse component $\gamma^{\mu}_{\perp}$. The physical interpretation of the process is very transparent in terms of the angular momentum transfer from the absorbed photon toward the incoming electron state: As the photon polarization is transverse by the Ward identity, its spin can only couple with the incoming electron spin in the transverse subspace.
In summary, the vector structure at tree-level exhibits a generalization of the $F_1$ form factor, that is now resolved into parallel and transverse components
\bea
\Gamma^{\mu}_{LO}(\overline{p},\overline{p}',l) &=& (2\pi)^3 \delta^{(3)}( \tilde{l} - \tilde{p} + \tilde{p}' ) \sum_{\lambda = \pm}\Delta^{\lambda}\left\{\gamma^{\mu}_{\parallel} F_{1,\parallel}^{\lambda,\lambda} + \gamma_{\perp}^{\mu} F_{1,\perp}^{\lambda,-\lambda} \right\},
\label{eq_F1gen}
\eea
with $F_{1,\parallel}^{\lambda,\lambda} = \mathcal{J}_{k_{\lambda},k'_{\lambda}}^{- -}(p^2,p'^2,l^1)$, and 
$F_{1,\perp}^{\lambda,-\lambda}  =\mathcal{J}_{k_{\lambda},k'_{-\lambda}}^{- -}(p^2,p'^2,l^1)$, respectively.
\subsection{Next to leading order contribution}\label{sec2B}

\begin{figure}[t!]
    \centering
    \includegraphics[width=0.5\linewidth]{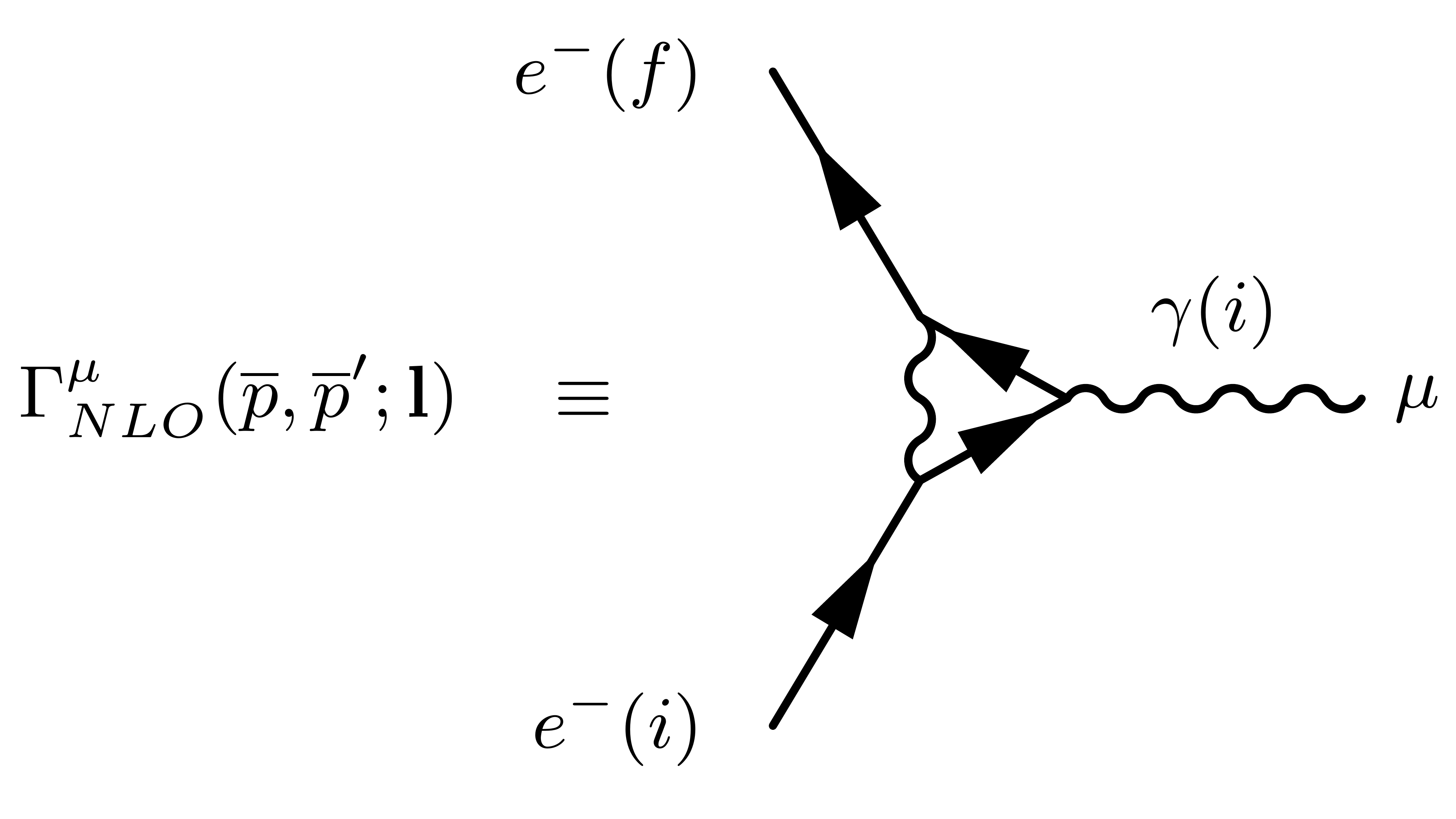}
    
    \caption{Feynman diagram for the next-to-leading order contribution to the fermion-photon vertex.}
    \label{fig3}
\end{figure}
At next to leading order (NLO), the corresponding amplitude is
\bea
\mathcal{V}_{NLO}(\breve{p},a,\breve{p}',a',\vec{l},c) &\!\!\!=\!\!\!& 
\langle e^{-}(\breve{p},a)|\ii^3 \int_{x,y,z}\mathcal{L}_{int}(x)\mathcal{L}_{int}(y)\mathcal{L}_{int}(z)|\gamma(\vec{l},c)e^{-}(\breve{p}',a')\rangle\nn\\
&\!\!\!=\!\!\!& \left(-\ii e\right)^3 \int_{x,y,z} \langle 0|b(\breve{p},a) \overline{\psi}(x)\gamma^{\nu}\psi(x)\overline{\psi}(y)\gamma^{\mu}\psi(y)\overline{\psi}(z)\gamma^{\alpha}\psi(z)A_{\nu}(x)A_{\mu}(y)A_{\alpha}(z)b^{\dagger}(\breve{p}',a')a^{\dagger}(\vec{l},c)|0\rangle.\nn\\
\eea
Details of the calculation can be found in Appendix~\ref{NLO}. After performing the contractions and inserting, as in the previous case, the Ritus representation of the eigenfunctions of the external states, we obtain
\bea
&&\mathcal{V}_{NLO}(\breve{p},a,\breve{p}',a',\vec{l},c) = \left(-\ii e\right)^3 \SumInt_{\left\{ \overline{q}_{e,1} \right\}}\sum_{a_1=1,2}\frac{1}{2 E_{e,1}} \SumInt_{\left\{ \overline{q}_{e,2} \right\}}\sum_{a_2=1,2}\frac{1}{2 E_{e,2}} \int_{l'_{E_{\gamma}}}\frac{1}{2 E'_{\gamma}}\sum_{c'}\epsilon_{\mu}(\vec{l}',c')\nn\\
&&\times\int_{x,y,x}\overline{U}(x,\overline{q}_1,a_1)\gamma^{\nu}
\ii S_F(x,y)\gamma^{\mu}\ii S_F(y,z) \gamma^{\alpha} \ii D_{\alpha\nu}(z,x)U(x,\overline{q}_2,a_2) e^{-\ii l'\cdot y}\nn\\
&&\times\langle 0| b(\breve{p},a) b^{\dagger}(\breve{q}_1,a_1) b(\breve{q}_2,a_2) b^{\dagger}(\breve{p}',a') |0\rangle \langle 0|a(\vec{l}',c') a^{\dagger}(\vec{l},c)|0\rangle .
\eea
Here, $S_F(x,y)$ and $D_{\alpha\nu}(z,x)$ are the Fermion and photon propagators, respectively.
The vacuum expectation values in the last factor are the same as in Eq.~\eqref{eq:bbbbaa1}, and hence the same result allows us to integrate the deltas in momenta and polarizations, to obtain
\bea
\mathcal{V}_{NLO}(\breve{p},a,\breve{p}',a',\vec{l},c) = -(\ii e)\, \epsilon_{\mu}(\vec{l},c)\overline{u}_e(k,p^3,a)\Gamma_{NLO}^{\mu}(\overline{p},\overline{p}',l) u_e (k',p'^{3},a'),
\eea
where the vertex contribution, at the next to leading order, is identified by the expression
\bea
\Gamma_{NLO}^{\mu}(\overline{p},\overline{p}',l) &=& \left(-\ii e\right)^2 \ii^3 \sum_{\lambda,\lambda'}(2\pi)^3 \delta^{(3)}( \tilde{l} - \tilde{p} + \tilde{p}' )\Delta^{\lambda}\mathcal{T}^{\mu}_{\lambda,\lambda'}(\overline{p},\overline{p}',l)\Delta^{\lambda'}.
\label{eq_GNLO}
\eea
Here, we defined
\bea
(2\pi)^3 \delta^{(3)}( \tilde{l} - \tilde{p} + \tilde{p}' )\mathcal{T}^{\mu}_{\lambda\lambda'}(\overline{p},\overline{p}',l) = \int_{x,y,z}\, \gamma^{\nu} S_F(x,y) \gamma^{\mu} S_F(y,z) \gamma^{\alpha} D_{\alpha\nu}(z,x) e^{-\ii l\cdot y} \mathcal{F}_e^*(x,\overline{p}_{\lambda}) \mathcal{F}_e(z,\overline{p}'_{\lambda'}).
\label{general}
\eea
Equation~(\ref{general}) represents the general expression describing  the fermion-photon vertex at one-loop level in the presence of the magnetic field. The induced tensor structure becomes rather rich. In this work, we explore in detail one of these structures  concentrating on the properties of the induced magnetic form factor in the purely transverse directions.  

\section{The $F_{2\perp\perp}^{\lambda\lambda'}$ form factor and the anomalous magnetic moment}\label{sec3}

Using the generalized Gordon identity, derived in Appendix~\ref{Gordon}, we identify the contribution to the vertex $i F_{2\perp\perp}^{\lambda\lambda'}\sigma_{\perp\perp}^{\mu 2}$ that leads to a generalized definition of the anomalous magnetic moment. Notice that this arises from the contraction of the one-loop correction to the vertex with external states, where according to the general spin and Minkowski space decomposition in Eq.~\eqref{eq_GNLO}, the tensor structures follow the general form
\begin{eqnarray}
\overline{u}_e(p_{\parallel},k,a)\Delta^{\lambda}\mathcal{T}^{\mu}_{\lambda,\lambda'}(\overline{p},\overline{p}',l)\Delta^{\lambda'}u_e(p'_{\parallel},k',a') = \overline{u}_e(p_{\parallel},k,a)\Delta^{\lambda}\left\{ (i F^{\lambda,\lambda}_{2\perp\perp}  + i F^{\lambda,-\lambda}_{2\perp\perp})\sigma_{\perp\perp}^{\mu 2} +  \ldots \right\}u_e(p'_{\parallel},k',a'),
\end{eqnarray}
where, to alleviate the notation, the condition $p^0 = E_e$ is from now on implicit in the spinor eigenfunctions $\left.u_e(p_{\parallel},k,a)\right|_{p^0 = E_e}$.

As shown in Appendix~\ref{NLO}, the explicit form of the factor $i F^{\lambda,\lambda'}_{2\perp\perp}(\overline{p},\overline{p}',l)$ reads
\begin{eqnarray}
i F^{\lambda,\lambda'}_{2\perp\perp}(\overline{p},\overline{p}',l) &=& -i\pi m \sqrt{2 B_e^3} e^{-\frac{l_{\perp}^2}{4 B_e}} e^{-\ii\frac{\left( p^2 + p^{'2} \right) l^1}{2 B_e}}
\sum_{n,n'}\sum_{\omega,\omega'}\sum_{\beta,\beta'}\sqrt{\frac{k_{\lambda}!n_{\omega'}!n'_{\beta'}!}{n_{\omega}!n'_{\beta}!k'_{\lambda'}!}}
e^{\ii\left( n_{\omega} - k_{\lambda} + n'_{\beta} - n_{\omega'} + k'_{\lambda'} - n'_{\beta'}  \right)\left(\varphi_{l} + \pi\right)}\nn\\
&\times&\Bigg\{ 
\delta_{\lambda',-\lambda}\delta_{\omega,\lambda}\delta_{\omega',\omega}\delta_{\omega,-\beta}\delta_{\beta',\beta}\left(\sqrt{k_{\lambda}} - \sqrt{k'_{\lambda'}}\right)+ \delta_{\omega,-\lambda}\delta_{\lambda',\lambda}\delta_{\omega',\omega}\delta_{\omega,-\beta}\delta_{\beta',-\beta}\sqrt{n'}
\nonumber\\ 
&&- \delta_{\omega,-\lambda}\delta_{\lambda',\lambda}\delta_{\omega',-\omega}\delta_{\omega,\beta}\delta_{\beta',\beta}\sqrt{n}
\Bigg\}\left( \frac{l_{\perp}^2}{2 B_e} \right)^{\frac{n'_{\beta} - n_{\omega'}}{2}} \mathcal{S}_{n_{\omega}+k'_{\lambda'}-k_{\lambda}-n'_{\beta'}} L_{n_{\omega'}}^{n'_{\beta} - n_{\omega'}}  \left( \frac{l_{\perp}^2}{2 B_e} \right)\nonumber\\
&\times& \int_{0}^{\infty}d\eta\,e^{-\eta}\,\eta^{\frac{n_{\omega} + k'_{\lambda'} - k_{\lambda} - n'_{\beta'}}{2}}
L_{k_{\lambda}}^{n_{\omega} - k_{\lambda}}\left( \eta\right) L_{n'_{\beta'}}^{k'_{\lambda'} - n'_{\beta'}}\left( \eta \right) J_{\left| n_{\omega} + k'_{\lambda'} - k_{\lambda} - n'_{\beta'}\right|}\left(  l_{\perp}\sqrt{\eta} \right) \mathcal{K}(\eta),
\label{eq_F2Full}
\end{eqnarray}
together with the integral over Feynman parameters 
\bea
\mathcal{K}(\eta) &=& \int_{x,y,z} \, \delta(x + y + z - 1) \cdot \frac{x}{\left[D\left( 2 B_e \eta \right)\right]^2}\nn\\
&=& \int \int \int dx\,dy\,dz \frac{x \delta(x + y + z - 1)}{\left[ \left( y p_{\parallel} + z p'_{\parallel} \right)^2 + 2\left( y n + z n'  \right) B_e + (y + z)m^2 - y p_{\parallel}^2 - z p_{\parallel}^{'2} + 2 B_e \eta x \right]^2},
\label{eq_Ieta}
\eea
which can be computed analytically, as shown in Appendix~\ref{Int_Feyn}. The {\rm{signature}} function is defined as $\mathcal{S}_m \equiv \left( \delta_{|m|,2 j } - \sgn(m)\delta_{|m|,2 j + 1 }\right)$. We notice that the magnetic field dependent factor $m B_e^{3/2}$ determines the dimensions of the form factor, since all the remaining terms are dimensionless. This is an explicit signature of the non-linear magnetic response of the system.

\subsection{The form factor $F_{2\perp\perp}^{\lambda\lambda'}$ in the limit $l\rightarrow 0$}

Let us now consider the limit of vanishing photon momentum (i.e. $l\rightarrow 0$) in the general expression, Eq.~\eqref{eq_F2Full}, for the form factor. We notice that in this limit, the expression vanishes unless the following conditions are met:
\begin{itemize}
\item The pre-factor $(l_{\perp}^2/2 B_e)^{(n'_{\beta} - n_{\omega'})/2}$ vanishes, except if
\begin{equation}
n'_{\beta} = n_{\omega'} \Longrightarrow n' - \frac{1 + \beta}{2} = n - \frac{1 + \omega'}{2}.
\label{eq_nn1}
\end{equation}
\item The Bessel function vanishes at the origin, since $J_{\nu}(z) \sim z^{\nu}$, except for $J_0 (z = 0) = 1$. Therefore, the only non-vanishing case is
\begin{equation}
n_{\omega} + k'_{\lambda'} - k_{\lambda} - n'_{\beta'} = 0.
\label{eq_summns}
\end{equation}
The upper limit for the upper index of the associated Laguerre polynomials $L_{n}^{\alpha}(z)$ is $\alpha > -1$. In particular, for $\alpha \in \mathbb{N}$ with $\alpha\le n$, the following relation holds
\bea
L_n^{-\alpha}(z) = \frac{(n-\alpha)!}{n!}(-z)^{\alpha} L_{n-\alpha}^{\alpha}(z),
\eea
thus indicating that $L_n^{-\alpha}(z) \sim (-z)^{\alpha}$ has a zero of order $\alpha$ at the origin. 
Therefore, we notice that Eq.~\eqref{eq_summns} only admits three possible sets of solutions:
\item Set 1: $n_{\omega} = k_{\lambda};\,\,\,\,\,k'_{\lambda'} = n'_{\beta'}$
\item Set 2: $n_{\omega} = n'_{\beta'};\,\,\,\,k'_{\lambda'} = k_{\lambda}$
\item Set 3: $n_{\omega} = -k'_{\lambda'};\,\,\,\,\,k_{\lambda} = - n'_{\beta'}$.

The Set 3 cannot be satisfied, since all four integers must me non-negative, otherwise the functions defined by Eq.~\eqref{eq_FQ} vanish.
\item To analyze if Set 2 can be fulfilled, we consider the following expression in Eq.~\eqref{eq_F2Full}
\bea
&&\Bigg\{\delta_{\lambda',-\lambda}\delta_{\omega,\lambda}\delta_{\omega',\omega}\delta_{\omega,-\beta}\delta_{\beta',\beta}\left(\sqrt{k_{\lambda}} - \sqrt{k'_{\lambda'}}\right)+ \delta_{\omega,-\lambda}\delta_{\lambda',\lambda}\delta_{\omega',\omega}\delta_{\omega,-\beta}\delta_{\beta',-\beta}\sqrt{n'}
\nonumber\\ 
&&- \delta_{\omega,-\lambda}\delta_{\lambda',\lambda}\delta_{\omega',-\omega}\delta_{\omega,\beta}\delta_{\beta',\beta}\sqrt{n}\Bigg\}
\label{eq_setterms}
\eea
If Set 2 is satisfied, then the first pair of terms in Eq.~\eqref{eq_setterms} exactly cancel. Considering now the Kronecker deltas in the third term, we notice that those imply $\beta' = -\beta$ and $\omega = \omega'$, and thus the condition $n_{\omega} = n'_{\beta}$ in Set 2 can equivalently be written as $n_{\omega'} = n'_{-\beta}$. The later, combined with Eq.~\eqref{eq_nn1}, would then imply $n'_{\beta} = n'_{-\beta}$, thus arriving at a contradiction since $\beta = \pm 1$. At last, considering the Kronecker deltas in the fourth term of Eq.~\eqref{eq_setterms}, those imply  $\beta'=\beta$ and $\omega=-\omega'$, and thus the condition $n_{\omega} = n'_{\beta'}$ can equivalently be expressed by $n_{-\omega'} = n'_{\beta}$. The later is again incompatible with  Eq.~\eqref{eq_nn1}, since it would imply $n_{\omega'} = n_{-\omega'}$, which cannot be satisfied for $\omega'=\pm1$. 

Therefore, we conclude that Set 2 cannot be satisfied, and we are only left with Set 1.
\end{itemize}

In summary, the sum is constrained by the following three conditions:
\begin{equation}
\label{eq_cond1}
n'_{\beta} = n_{\omega'} \Longrightarrow n' - \frac{1 + \beta}{2} = n - \frac{1 + \omega'}{2},
\end{equation}
\begin{equation}
\label{eq_cond2}
n_{\omega} = k_{\lambda} \Longrightarrow n - \frac{1+\omega}{2} = k - \frac{1+\lambda}{2},\\
\end{equation}
\begin{equation}
n'_{\beta'} = k'_{\lambda'} \Longrightarrow n' - \frac{1+\beta'}{2} = k' - \frac{1 + \lambda'}{2}.
\label{eq_cond3}
\end{equation}

Notice that, under the system of conditions in Eqs.~\eqref{eq_cond1},~\eqref{eq_cond2} and~\eqref{eq_cond3}, the ratio of factorials in front of Eq.~\eqref{eq_F2Full} cancels,
\begin{eqnarray}
\sqrt{\frac{k_{\lambda}!n_{\omega'}!n'_{\beta'}!}{n_{\omega}!n'_{\beta}!k'_{\lambda'}!}} \rightarrow1,
\end{eqnarray}
and also the phase factor becomes equal to 1. Moreover, the signature
\begin{eqnarray}
\mathcal{S}_{n_{\omega}+k'_{\lambda'}-k_{\lambda}-n'_{\beta'}} \rightarrow \mathcal{S}_{0} = 1.
\end{eqnarray}
Finally, the Laguerre polynomial
\begin{eqnarray}
\lim_{l_{\perp}\rightarrow 0} L_{n_{\omega'}}^{n'_{\beta} - n_{\omega'}}\left(  \frac{l_{\perp}^2}{2 B_e}\right) = L_{n_{\omega'}}^{n'_{\beta} - n_{\omega'}}\left( 0\right) = 1.
\end{eqnarray}
The expression can be further simplified by solving for $n$ and $n'$ in favor of the external indexes as follows:
From Eq.~\eqref{eq_cond2} and Eq.~\eqref{eq_cond3}, we have
\begin{eqnarray}
n &=& n_{\omega} + \frac{1 + \omega}{2} = k_{\lambda} + \frac{1+\omega}{2}\nonumber\\
n' &=& n'_{\beta'} + \frac{1 + \beta'}{2} = k'_{\lambda'} +  \frac{1 + \beta'}{2}.
\end{eqnarray}
Furthermore, in the second set of terms in Eq.~\eqref{eq_F2Full}, the $\sqrt{n'}$ goes with a product of Kronecker deltas that implies the condition $\omega = -\lambda = -\lambda' = \beta'$, and hence there we can substitute $\sqrt{n'} \rightarrow \sqrt{k'_{\lambda'} + \frac{1-\lambda}{2}}$. Similarly, for the third term, the $\sqrt{n}$ goes with a product of Kronecker deltas that implies $\omega = -\lambda = -\lambda'$. Therefore, in this second term we may substitute $\sqrt{n}\rightarrow \sqrt{k_{\lambda}+\frac{1-\lambda}{2}}$. Summarizing the previous analysis, we arrive at the expression
\begin{eqnarray}
i F^{\lambda,\lambda'}_{2\perp\perp}(p,p',l=0) &\!\!=\!\!&  -i\pi m \sqrt{2 B_e^3} 
\sum_{n,n'}\sum_{\omega,\omega'}\sum_{\beta,\beta'} \delta_{n'_{\beta},n_{\omega'}}\delta_{n_{\omega},k_{\lambda}}\delta_{n'_{\beta'},k'_{\lambda'}}\Bigg\{ 
\delta_{\lambda',-\lambda}\delta_{\omega,\lambda}\delta_{\omega',\omega}\delta_{\omega,-\beta}\delta_{\beta',\beta}\left(\sqrt{k_{\lambda}} - \sqrt{k'_{\lambda'}}\right)\nonumber\\
&\!\!+\!\!& \delta_{\lambda',\lambda}\delta_{\omega,-\lambda}\left(\delta_{\omega',\omega}\delta_{\omega,-\beta}\delta_{\beta',-\beta}\sqrt{k'_{\lambda'} +\frac{1-\lambda}{2}}
- \delta_{\omega',-\omega}\delta_{\omega,\beta}\delta_{\beta',\beta}\sqrt{k_{\lambda} +\frac{1-\lambda}{2}}\right)
\Bigg\}\mathcal{I}(k_{\lambda},k'_{\lambda'}),
\label{eq_F2l02-1}
\end{eqnarray}
where the conditions Eq.~\eqref{eq_cond1},~\eqref{eq_cond2},~and Eq.~\eqref{eq_cond3} are summarized as Kronecker deltas in the sum. We also defined, to alleviate the notation, the function
\begin{equation}
\mathcal{I}(k_{\lambda},k'_{\lambda'}) = \int_{0}^{\infty}d\eta\,e^{-\eta}\,
L_{k_{\lambda}}\left( \eta\right) L_{k'_{\lambda'}}\left( \eta \right) \mathcal{K}(\eta),
\label{eq_Ikkp}
\end{equation}
where explicit analytical expressions for the function $\mathcal{K}(\eta)$ for different cases are presented in Appendix~\ref{Int_Feyn}, while the integral in Eq.~\eqref{eq_Ikkp} is evaluated numerically.  We now proceed to study the selection rules for transitions between different fermion Landau levels.
\subsection{Selection rules}
\begin{enumerate}
\item $k=k'=0$: In this case $\lambda = \lambda' = -1$ (LLL to LLL). Therefore, Eq.~(\ref{eq_F2l02-1}) vanishes, and the process is forbidden. This verifies the result found in Ref.~\cite{Fraga:2024klm}.

\item $k=1$, $k'=0$ (1LL to LLL): For this case we can have two possible values for $k_\lambda$ and a single value for $k'_{\lambda'}$
\begin{eqnarray}
k_\lambda&=&k -\left(\frac{1+\lambda}{2}\right)=\left\{
\begin{array}{cl}
   0  & (\lambda=1) \\
   1  & (\lambda=-1)
\end{array}\right.\nonumber\\
k'_{\lambda'}&=&0\ \ (\lambda'=-1).
\end{eqnarray}
For the first term in Eq.~(\ref{eq_F2l02-1}), since $\lambda'=-\lambda$ and there's only one possible value for $\lambda'$ then, both $k_\lambda=k'_{\lambda'}=0$ and this term vanishes. The second term in Eq.~(\ref{eq_F2l02-1}) contributes with a factor $\sqrt{k'_{\lambda'} + (1-\lambda)/2} = 1$, since $\lambda'=\lambda = -1$. Finally, the third term contributes with a factor $-\sqrt{k_{\lambda} + (1 - \lambda)/2} = -\sqrt{2}$. Therefore

\begin{equation}
i F^{\lambda,\lambda'}_{2\perp\perp}(k = 1  ,k'=0,l\rightarrow 0) =  -i\pi m \sqrt{2 B_e^3}
\,\delta_{\lambda,-1}\delta_{\lambda',-1}\left( 1 - \sqrt{2} \right)\mathcal{I}(1,0).
\end{equation}

\item $k=0$, $k'=1$ (LLL to 1LL): For this case we can have two possible values for $k'_{\lambda'}$ and a single value for $k_{\lambda}$
\begin{eqnarray}
k'_{\lambda'}&=&k' -\left(\frac{1+\lambda'}{2}\right)=\left\{
\begin{array}{cl}
   0  & (\lambda'=1) \\
   1  & (\lambda'=-1)
\end{array}\right.\nonumber\\
k_\lambda&=&0\ \ (\lambda=-1).
\end{eqnarray}
For the first term in Eq.~(\ref{eq_F2l02-1}), since $\lambda=-\lambda'$, we have $k_\lambda=k'_{\lambda'}=0$ and this term vanishes.
The second term in Eq.~(\ref{eq_F2l02-1}) contributes with a factor $\sqrt{k'_{\lambda'} + (1-\lambda)/2} = \sqrt{2}$, since $\lambda'=\lambda = -1$. Finally, the third term contributes with a factor $-\sqrt{k_{\lambda} + (1 - \lambda)/2} = -1$. Therefore
\begin{equation}
i F^{\lambda,\lambda'}_{2\perp\perp}(k = 0  ,k'=1,l\rightarrow 0) =  i\pi m \sqrt{2 B_e^3}
\,\delta_{\lambda,-1}\delta_{\lambda',-1}\left( 1 - \sqrt{2} \right)\mathcal{I}(1,0).
\end{equation}


Notice that the magnetic form factor $ F_{2\perp\perp}(k = 0  ,k'=1,l\rightarrow 0)=-F_{2\perp\perp}(k = 1  ,k'=0,l\rightarrow 0)$. This reflects the loss of time reversal invariance due to the presence of the magnetic field.

\end{enumerate}

In a similar fashion, one can show that the magnetic form factors for transitions LLL to 2LL, 2LL to LLL and 1LL to 1LL are given by

\begin{enumerate}

\item[4.] $k=0$, $k'=2$ (LLL to 2LL):
\begin{equation}
i F^{\lambda,\lambda'}_{2\perp\perp}(k=0,k'=2,l\to 0)=-i\pi m \sqrt{2 B_e^3}\left\{  -\delta_{\lambda,-1}\delta_{\lambda',1}\mathcal{I}(0,1)
+ \delta_{\lambda,-1}\delta_{\lambda',-1}\left( \sqrt{3} - 1 \right)\mathcal{I}(0,2)
\right\}.
\end{equation}
\item[5.] $k=2$, $k'=0$ (2LL to LLL): 
\begin{equation}
i F^{\lambda,\lambda'}_{2\perp\perp}(k=2,k'=0,l\to 0)=i\pi m \sqrt{2 B_e^3}\left\{  -\delta_{\lambda,-1}\delta_{\lambda',1}\mathcal{I}(0,1)
+ \delta_{\lambda,-1}\delta_{\lambda',-1}\left( \sqrt{3} - 1 \right)\mathcal{I}(0,2)
\right\},
\end{equation}
showing that once again these two form factors are opposite in sign, reflecting the loss of time reversal invariance.

\item[6.] $k=1$, $k'=1$ (1LL to 1LL):
\begin{eqnarray}
i F^{\lambda,\lambda'}_{2\perp\perp}(k=1,k'=1,l\to 0)&=&-i\pi m \sqrt{2 B_e^3}\Big\{-\delta_{\lambda,1}\delta_{\lambda',-1}\mathcal{I}(0,1)
+\delta_{\lambda,-1}\delta_{\lambda',1}\mathcal{I}(1,0)\nonumber\\ 
&+& \delta_{\lambda,1}\delta_{\lambda',1}\sqrt{2}\mathcal{I}(1,1)
-\delta_{\lambda,-1}\delta_{\lambda',-1}\sqrt{2}\mathcal{I}(1,1)
\Big\}.
\end{eqnarray}

\end{enumerate}

\begin{figure}[t!]
    \centering
    \includegraphics[width=0.5\linewidth]{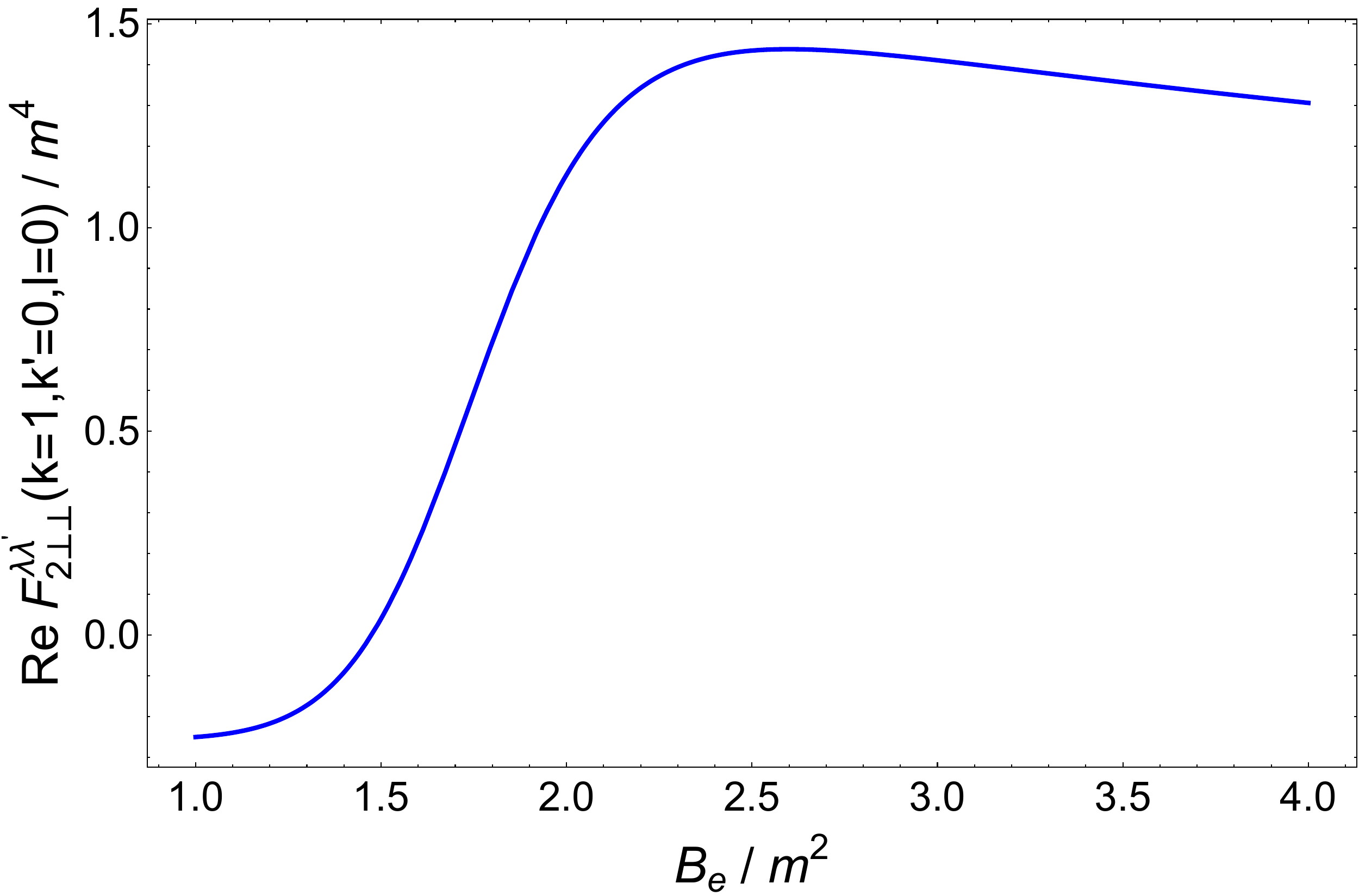}
    
    \caption{Real part of the magnetic form factor, normalized to $m^4$, as a function of the background magnetic field intensity, normalized to $m^2$, for the transition $k = 1 \rightarrow k' = 0$ between the initial and final fermion states that scatter with the photon. }
    \label{real}
\end{figure}

\begin{figure}
    \centering
    \includegraphics[width=0.5\linewidth]{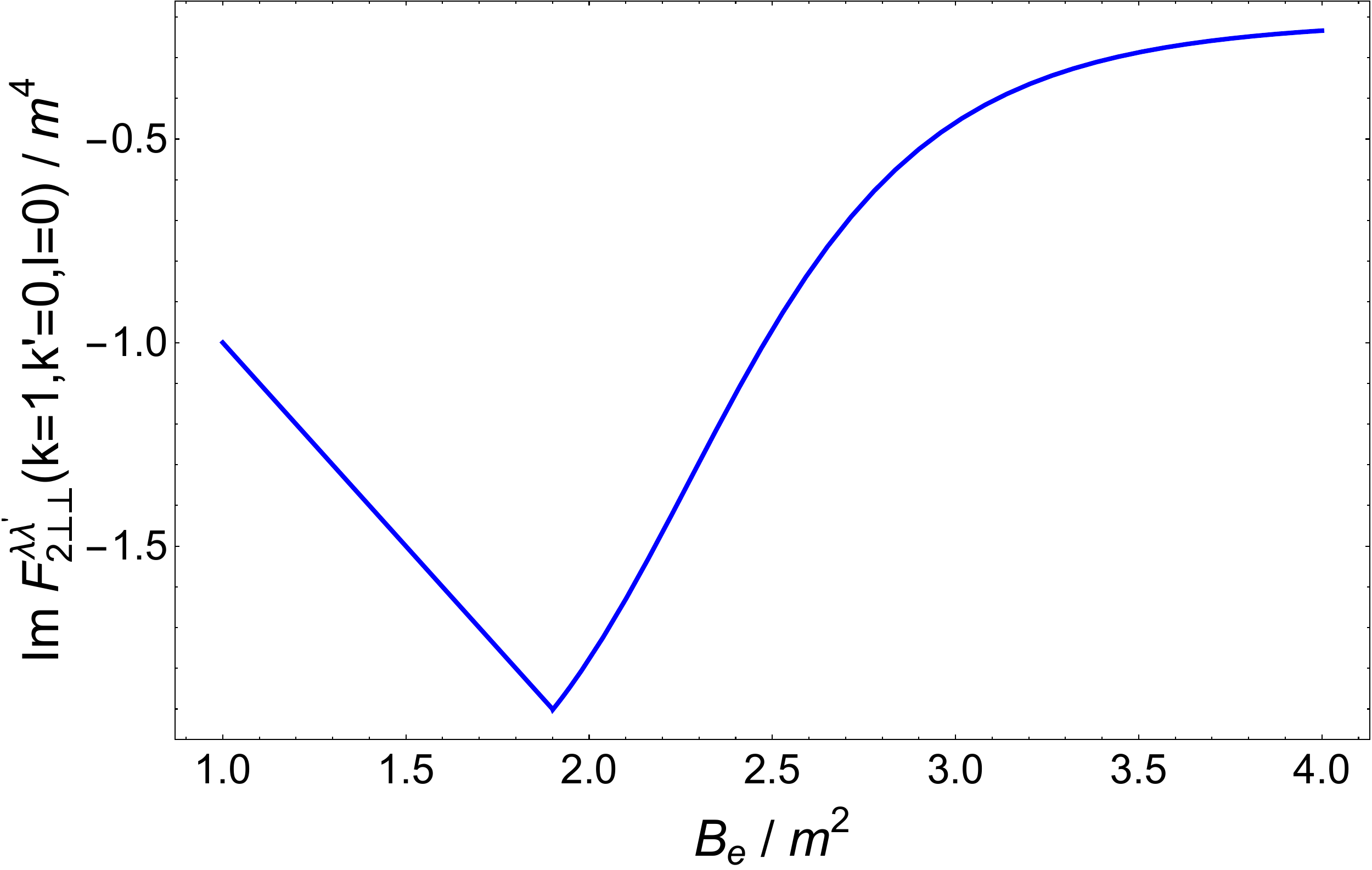}
    \caption{Imaginary part of the magnetic form factor, normalized to $m^4$, as a function of the background magnetic field intensity, normalized to $m^2$. This is interpreted as being related to a finite lifetime for the transition $k = 1 \rightarrow k' = 0$ between the initial and final fermion states that scatter with the photon. }
    \label{imag}
\end{figure}

As an example of the behavior of the AMM form factor, we compute $F^{\lambda,\lambda'}_{2\perp\perp}(k = 1  ,k'=0,l\rightarrow 0)$. Figure~\ref{real} and~\ref{imag} show the real and imaginary parts of this amplitude, respectively, normalized to $m^4$, as a function of the magnetic field normalized to $m^2$. Notice that the real part corresponds to a non-linear field-dependent AMM that exhibits saturation behavior in the very strong magnetic field limit $|B_e|/m^2\gg 1$. We emphasize that this behavior is reminiscent of the magnetic response in ferromagnetic media, where the magnetization is a non-linear function of the external magnetic field due to microscopic spin polarization. In this sense, the Ritus eigenbasis, where the spin polarization is a good quantum number, provides the correct physical picture for this magnetization mechanism.
In contrast, the imaginary part of the form factor represents a finite lifetime (or decay rate) for the fermion excited states that emerge from the scattering process.

\section{Summary and Conclusions}\label{concl}

In this work, we have computed the magnetic field modifications to the fermion-photon vertex in QED up to one-loop order for arbitrary field strengths. We have explicitly shown that even at tree level, this vertex is modified due to the breaking of Lorentz symmetry, which results in the vertex splitting into longitudinal and transverse pieces. A remarkable feature of the emerging structure at tree-level is that the spin polarization of the scattered electron state is preserved for the parallel vector component, while it reverses for the transverse component. This is due to the  angular momentum transfer from the photon to the fermion: since the photon polarization is transverse by gauge invariance, its spin can only couple to the incoming electron spin in the transverse subspace. The one-loop modifications induce a rather rich tensor structure for the vertex; in particular, the emergence of terms that correspond to the magnetic field-induced modifications of the AMM. Once again, due to the breaking of Lorentz symmetry, the AMM receives contributions that are purely longitudinal, purely transverse, and also mixed longitudinal/transverse. In this work, we have concentrated on describing in detail the purely transverse contribution to the AMM and have found the selection rules for a few of the lowest lying transitions for fermions occupying different Landau levels. The transition amplitudes describing the purely transverse AMM form factor are, in general, complex, and we interpret the phase factor as being related to the life-time of the initial state that makes a transition to a given final state. The amplitude for transitions from an initial to a final Landau level, differ by a sign from the reverse process due to the loss of time reversal invariance induced by the presence of the field. We have also shown that for the purely transverse AMM, transitions from LLL to LLL are forbidden, in agreement with the findings of Ref.~\cite{Fraga:2024klm}. Moreover, transitions  between Landau levels other than the LLL are infrared finite; thus, unlike the findings of Ref.~\cite{Fraga:2024klm}, the introduction of a gluon/photon mass is unnecessary since the magnetic field provides a natural infrared cut-off. The rich tensor structure that we found is a subject that requires further exploration. This is work in progress that will be reported elsewhere.

\section*{Acknowledgments}
E.M. acknowledges financial support from ANID Fondecyt Grant No. 1230440. A.A. acknowledges support from DGAPA-PAPIIT-UNAM grant number IG100826. and from Secretar\'\i a de Ciencia, Humanidades, Tecnolog\'\i a e Innovaci\'on (SECIHTI) M\'exico grant number CIORGANISMOS-2025-17.
ANID Fondecyt Grants No. 1260183, No. 1250206, and No. 1241436.

\appendix

\section{The Vertex at the tree-level (leading order)}
\label{LO}
At the leading order (LO), the $e^{-} + \gamma\rightarrow e^{-}$, 
the corresponding amplitude is given by the expression
\bea
\mathcal{V}_{LO}(\breve{p},a,\breve{p}',a';\vec{l},c) &=& \langle e^{-}(\breve{p},a)|\ii \int_{x}\mathcal{L}_{int}(x)|\gamma(\vec{l},c)e^{-}(\breve{p}',a')\rangle\nn\\
&=& -\ii e \int d^4 x \langle e^{-}(\breve{p},a)| \overline{\psi}(x)\gamma^{\mu}\psi(x)A_{\mu}(x)|\gamma(\vec{l},c)e^{-}(\breve{p}',a')\rangle\nn\\
&=& -\ii e \int d^4 x \langle 0|b(\breve{p},a) \overline{\psi}(x)\gamma^{\mu}\psi(x)A_{\mu}(x)b^{\dagger}(\breve{p}',a')a^{\dagger}(\vec{l},c)|0\rangle
\label{eq:VLO}
\eea
At this point, we need to consider the expansion of the spinors according to Eq.~\eqref{eq:spinor}, 
\bea
\overline{\psi}(x) &=& \SumInt_{\{ \overline{q}_{E_{e,1}} \}} \sum_{a_1=1,2}\frac{1}{2 E_{e,1}}\Bigg\{ b^{\dagger}(\breve{q}_1,a_1) \overline{U}(x,\overline{q}_1,a_1)+ d(\breve{q}_1,a_1) \overline{V}(x,\overline{q}_1,a_1)
\Bigg\}\nn\\
\psi(x) &=& \SumInt_{\{ \overline{q}_{E_{e,2}} \}} \sum_{a_2=1,2}\frac{1}{2 E_{e,2}}\Bigg\{ b(\breve{q}_2,a_2) U(x,\overline{q}_2,a_2)+ d^{\dagger}(\breve{q}_2,a_2) V(x,\overline{q}_2,a_2)
\Bigg\},
\eea
as well as the representation of the gauge field
\bea
A^{\mu}(x) = \int_{l'_{E_{\gamma}}}\frac{1}{2 E'_{\gamma}}\sum_{c'}\left\{
a(\vec{l}',c')\epsilon^{\mu}(\vec{l'},c')e^{-\ii l'\cdot x}+a^{\dagger}(\vec{l},c)\left[ \epsilon^{\mu}(\vec{l'},c')\right]^* e^{\ii l'\cdot x}
\right\}.
\eea
Inserting these into Eq.~\eqref{eq:VLO}, we recognize that the nonzero terms are the following
\bea
\mathcal{V}_{LO}(\breve{p},a,\breve{p}',a',\vec{l},c) &\!\!=\!\!& -\ii e \int d^4 x \SumInt_{\{ \overline{q}_{E_{e,1}} \}} \sum_{a_1=1,2}\frac{1}{2 E_{e,1}} \SumInt_{\{ \overline{q}_{E_{e,2}} \}} \sum_{a_2=1,2}\frac{1}{2 E_{e,2}} \int_{l'_{E_{\gamma}}}\frac{1}{2 E'_{\gamma}}\sum_{c'}\epsilon_{\mu}(\vec{l'},c') \overline{U}(x,\overline{q}_1,a_1)\gamma^{\mu}U(x,\overline{q}_2,a_2)\nn\\
&\!\!\times\!\!&  e^{-\ii l'\cdot x}\langle 0| b(\breve{p},a)b^{\dagger}(\breve{q}_1,a_1) b(\breve{q}_2,a_2)b^{\dagger}(\breve{p}',a')   |0\rangle
\langle 0| a(\vec{l}',c')a^{\dagger}(\vec{l},c) |0\rangle.
\eea
Here, we calculate the matrix elements applying Wick's theorem, to obtain
\bea
\langle 0| a(\vec{l}',c')a^{\dagger}(\vec{l},c) |0\rangle &=& - g^{cc'}2 E_{\gamma}(2\pi)^3\delta^{(3)}(\vec{l}'-\vec{l})\nn\\
\langle 0| b(\breve{p},a)b^{\dagger}(\breve{q}_1,a_1) b(\breve{q}_2,a_2)b^{\dagger}(\breve{p}',a')   |0\rangle
&=& 2 E_{e,1}\delta_{a_1,a}(2\pi)^{2}\delta_{k_1,k}\delta(q_1^2 - p^2)\delta(q_1^3 - p^3)\nn\\
&&\times 2 E_{e,2}\delta_{a_2,a'}(2\pi)^{2}\delta_{k_2,k'}\delta(q_2^2 - p'^2)\delta(q_2^3 - p'^3).
\label{eq:bbbbaa}
\eea
Integrating out the delta functions, we end up with the expression
\bea
\mathcal{V}_{LO}(\breve{p},a,\breve{p}',a',\vec{l},c) = -\ii e\, \epsilon_{\mu}(\vec{l},c)\int d^4 x \overline{U}(x,\overline{p},a)\gamma^{\mu}U(x,\overline{p}',a') e^{-\ii l\cdot x}.
\eea
Now, we insert the definition of the spinor amplitudes in the Ritus basis according to Eq.~\eqref{eq_URit},
\bea
U(x,\overline{p}',a') &=& \mathbb{E}^e(x,\overline{p}') u_e (p'_{\parallel},k',a') = \sum_{\lambda' = \pm}\Delta^{\lambda'}\mathcal{F}_e(x,\overline{p}'_{\lambda'})u_e (p'_{\parallel},k',a')\nn\\
\overline{U}(x,\overline{p},a) &=& \mathbb{E}^{e\dagger} (x,\overline{p}) \overline{u}_e (p_{\parallel},k,a) = \sum_{\lambda = \pm}\Delta^{\lambda}\mathcal{F}_e^{*}(x,\overline{p}_{\lambda})\overline{u}_e (p_{\parallel},k,a)
\eea
such that the amplitude reduces to the form
\bea
\mathcal{V}_{LO}(\breve{p},a,\breve{p}',a',\vec{l},c) = -\ii e\, \epsilon_{\mu}(\vec{l},c) \overline{u}_e (p_{\parallel},k,a) \Gamma^{\mu}_{LO}(\overline{p},\overline{p}',l)u_e (p'_{\parallel},k',a')
\eea
where we identify the expression for the vertex at the tree-level
\bea
\Gamma^{\mu}_{LO}(\overline{p},\overline{p}',l) = \sum_{\lambda,\lambda'}\Delta^{\lambda}\gamma^{\mu}\Delta^{\lambda'}\int d^4x \mathcal{F}_e^{*}(x,\overline{p}_{\lambda})\mathcal{F}_e(x,\overline{p}'_{\lambda'})e^{-\ii l\cdot x}
\eea
Now, we apply the identity
\bea
\Delta^{\lambda}\gamma^{\mu}\Delta^{\lambda'} = \Delta^{\lambda}\left[ \delta_{\lambda',\lambda}\gamma_{\parallel}^{\mu}
+ \delta_{\lambda',-\lambda}\gamma_{\perp}^{\mu}\right]
\eea
to obtain
\bea
\Gamma^{\mu}_{LO}(\overline{p},\overline{p}',l) = \sum_{\lambda}\Delta^{\lambda}\left\{\gamma^{\mu}_{\parallel}\int d^4x \mathcal{F}_e^{*}(x,\overline{p}_{\lambda})\mathcal{F}_e(x,\overline{p}'_{\lambda})e^{-\ii l\cdot x} + \gamma_{\perp}^{\mu} \int d^4x \mathcal{F}_e^{*}(x,\overline{p}_{\lambda})\mathcal{F}_e(x,\overline{p}'_{-\lambda})e^{-\ii l\cdot x}\right\}
\eea
Finally, we can explicitly evaluate the integrals as follows (we pick $s = s' = -1$ for electrons in QED)
\bea
\int d^4x \mathcal{F}_e^{*}(x,\overline{p}_{\lambda})\mathcal{F}_e(x,\overline{p}'_{\lambda'})e^{-\ii l\cdot x} &=& \int dx^0 e^{-\ii(l^0 - p^0 + p'^0)}\int dx^2 e^{\ii(l^2 - p^2 + p'^2)}\int dx^3 e^{\ii (l^3 - p^3 + p'^3)x^3}\nn\\
&&\times N_{k_{\lambda}}N_{k'_{\lambda'}} \int_{-\infty}^{\infty}dx^1 D_{k_{\lambda}}\left( \sqrt{2 B_e}(x^1 + \frac{p^2}{B_e}) \right) D_{k'_{\lambda'}}\left( \sqrt{2 B_e}( x^1 + \frac{p'^2}{B_e} ) \right) e^{\ii l^1 x^1}\nn\\
&=& (2\pi)^3 \delta( l^0 - p^0 + p'^0 )\delta( l^2 - p^2 + p'^2 )\delta( l^3 - p^3 + p'^3 ) \mathcal{J}_{k_{\lambda},k'_{\lambda'}}^{s s}(p^2,p'^2,l^1)
\eea
As shown in App.~\ref{IntegralsDD}, the last factor is defined by
\bea
\mathcal{J}_{k_{\lambda},k'_{\lambda'}}^{s s}(p^2,p'^2,l^1) &=& e^{\ii \frac{(s p^2 + s p'^2)l^1}{\sqrt{2 B_e}}} \mathcal{G}_{k_{\lambda},k'_{\lambda'}}^{s s}(l^1,l^2 = s p'^2 - s p^2)\nn\\
&=& 2\pi e^{\ii \frac{(s p^2 + s p'^2)l^1}{\sqrt{2 B_e}}} e^{-\frac{l_{\perp}^2}{4 B_e}} \sqrt{\frac{k_{s\lambda}!}{k'_{s\lambda'}!}}\ii^{k'_{s\lambda'}-k_{s\lambda}}e^{\ii(k'_{s\lambda'} - k_{s\lambda})\phi_{\perp}}\left( \frac{l_{\perp}^2}{2 B_e} \right)^{\frac{k'_{s\lambda'}-k_{s\lambda}}{2}}  L_{k_{s\lambda}}^{k'_{s\lambda'}-k_{s\lambda}}\left( \frac{l_{\perp}^2}{2 B_e} \right),
\eea
where
\bea
\tan\phi_{\perp} = \frac{s p'^2 - s p^2}{l^1},
\eea
and $l_{\perp} = (l^1,l^2)$.
Therefore, our final result for the vertex at tree-level is
\bea
\Gamma^{\mu}_{LO}(\overline{p},\overline{p}',l) &=& (2\pi)^3 \delta( l^0 - p^0 + p'^0 )\delta( l^2 - p^2 + p'^2 )\delta( l^3 - p^3 + p'^3 ) \nonumber\\
&\times&
\sum_{\lambda}\Delta^{\lambda}\left\{\gamma^{\mu}_{\parallel} \mathcal{J}_{k_{\lambda},k'_{\lambda}}^{- -}(p^2,p'^2,l^1) + \gamma_{\perp}^{\mu} \mathcal{J}_{k_{\lambda},k'_{-\lambda}}^{- -}(p^2,p'^2,l^1) \right\}\nn\\
\eea


\section{The integrals involved in the calculation of the vertex at tree-level (LO)}
\label{IntegralsDD}
The generic integrals involved in the calculation of the vertex functions introduced in Eq.~\eqref{eq_Jfunc} are of the form
\bea
\mathcal{J}_{k k'}^{s s'}(p^2,p'^2,l^1) = N_k N_{k'}\int_{-\infty}^{\infty}dx^1 e^{\ii l^1 x^1 }D_k\left( \sqrt{2 B_e}x^1 -s\sqrt{\frac{2}{B_e}}p^2 \right) D_{k'}\left( \sqrt{2 B_e}x^1 -s'\sqrt{\frac{2}{B_e}}p'^2 \right)
\label{eq:Jkk}
\eea
In order to compute this integral, we shall use the identity (Gradshteyn, p.838, Eq.7.377), valid for $n\ge m$,
\bea
\int_{-\infty}^{\infty}e^{-x^2}H_m(x + y)H_n(x+z)dx= 2^n \pi^{1/2}m! z^{n-m} L_m^{n-m}(-2yz)
\label{eq:idgrad2}
\eea
For this purpose, let us start by considering the more general expression
\begin{widetext}
\bea
\mathcal{J} &=& \int_{-\infty}^{\infty}d\chi e^{\ii \alpha \chi}D_{m}(\chi + \kappa)D_{n}(\chi - \kappa)= 2^{-\frac{(n+m)}{2}} \int_{-\infty}^{\infty}d\chi\, e^{\ii\alpha\chi - \frac{(\chi+\kappa)^2}{4}-\frac{(\chi-\kappa)^2}{4}} H_{m}\left( \frac{\chi + \kappa}{\sqrt{2}} \right) H_{n}\left( \frac{\chi - \kappa}{\sqrt{2}} \right)\nn\\
&=& 2^{-\frac{(n+m)}{2}} e^{-\frac{\alpha^2 + \kappa^2}{2}} \int_{-\infty}^{\infty}d\chi\, e^{ - \frac{(\chi-\ii\alpha)^2}{2}} H_{m}\left( \frac{\chi + \kappa}{\sqrt{2}} \right) H_{n}\left( \frac{\chi - \kappa}{\sqrt{2}} \right).
\eea
\end{widetext}
Let us define the variables
\bea
x &=& \frac{\chi - \ii\alpha}{\sqrt{2}} \Longrightarrow \chi = \sqrt{2}x + \ii\alpha,
\eea
such that the combinations in the arguments of the Hermite polynomials become
\bea
\frac{\chi + \kappa}{\sqrt{2}} &=& x + \frac{\kappa + \ii\alpha}{\sqrt{2}}\nn\\
\frac{\chi - \kappa}{\sqrt{2}} &=& x - \frac{\kappa - \ii\alpha}{\sqrt{2}},
\eea
and hence upon substituting we have, applying the identity in Eq.~\eqref{eq:idgrad2} for the case $n\ge m$
\begin{widetext}
\bea
\mathcal{J} &=& \sqrt{2}\,2^{-\frac{(n+m)}{2}} e^{-\frac{\alpha^2 + \kappa^2}{2}} \int_{-\infty}^{\infty}dx e^{-x^2} H_m\left( x + \frac{\kappa + \ii\alpha}{\sqrt{2}}  \right) H_n\left( x - \frac{\kappa - \ii\alpha}{\sqrt{2}}\right)\nn\\
&=& \sqrt{2}\,2^{-\frac{(n+m)}{2}} e^{-\frac{\alpha^2 + \kappa^2}{2}} 2^n\pi^{1/2}m!\left( - \frac{\kappa - \ii\alpha}{\sqrt{2}} \right)^{n-m}L_m^{n-m}\left( 2\frac{\left( \kappa - \ii\alpha \right)\left( \kappa + \ii\alpha \right)}{2} \right)\nn\\
&=& \sqrt{2\pi}e^{-\frac{\alpha^2 + \kappa^2}{2}} m! \left( \ii\alpha - \kappa \right)^{n-m}L_m^{n-m}\left( \kappa^2 + \alpha^2 \right),\,\,\,\,n\ge m
\eea
\end{widetext}
Reversing the argument for the case $n\le m$, we obtain a similar expression. Therefore, we can summarize the result as
\begin{widetext}
\bea
\mathcal{J} = \int_{-\infty}^{\infty}d\chi e^{\ii \alpha \chi}D_{m}(\chi + \kappa)D_{n}(\chi - \kappa) = \left\{ 
 \begin{array}{cc}
\sqrt{2\pi}e^{-\frac{\alpha^2 + \kappa^2}{2}} m! \left( \ii\alpha - \kappa \right)^{n-m}L_m^{n-m}\left( \kappa^2 + \alpha^2 \right), & n\ge m\\
\sqrt{2\pi}e^{-\frac{\alpha^2 + \kappa^2}{2}} n! \left( \ii\alpha + \kappa \right)^{m-n}L_n^{m-n}\left( \kappa^2 + \alpha^2 \right), & m\ge n
 \end{array}
 \right.
 \label{eq:J1}
\eea
\end{widetext}
Let us now go back to the general integral expression defined in Eq.~\eqref{eq:Jkk}, to apply the previous result we define
\bea
\chi + \kappa &=& \sqrt{2 B_e}x^{1} - s\sqrt{\frac{2}{Be}}p^2\nn\\
\chi - \kappa &=& \sqrt{2 B_e}x^{1} - s'\sqrt{\frac{2}{Be}}p'^2.
\eea
Combining these two, we have
\bea
\chi &=& \sqrt{2 B_e}x^1 - \frac{1}{\sqrt{2 Be}}\left(  s p^2 + s' p'^2 \right)\nn\\
\kappa &=& \frac{1}{\sqrt{2Be}} \left(  s' p'^2 - s p^2 \right),
\eea
with $dx^1 = d\chi /\sqrt{2 B_e}$. Substituting this change of variables into Eq.~\eqref{eq:Jkk}, with
$\alpha=l^{1}/\sqrt{2 B_e}$ as follows
\begin{widetext}
\bea
\mathcal{J}_{k k'}^{s s'}(p^2,p'^2,l^1) &=& \frac{N_k N_{k'}}{\sqrt{2 B_e}}\int_{-\infty}^{\infty}d\chi e^{\ii \frac{l^1}{\sqrt{2 B_e}} \chi }D_k\left( \chi + \kappa\right) D_{k'}\left( \chi - \kappa \right)\nn\\ 
&=& 2\pi e^{\ii\frac{(s p^2 + s' p'^2)l^1}{2 B_e}} e^{-\frac{(l^1)^2 + (s' p'2 - s p^2)^2}{4 B_e}}\left\{
\begin{array}{cc}
\sqrt{\frac{k!}{k'!}}\left(\frac{ \ii l^1 - (s' p'^2 - s p^2) }{\sqrt{2 B_e}}\right)^{k' - k} L_{k}^{k'-k}\left( \frac{(l^1)^2 + (s' p'^2 - s p^2)^2}{2 B_e} \right) & k'\ge k\\
\sqrt{\frac{k'!}{k!}}\left(\frac{ \ii l^1 + (s' p'^2 - s p^2) }{\sqrt{2 B_e}}\right)^{k - k'} L_{k'}^{k-k'}\left( \frac{(l^1)^2 + (s' p'^2 - s p^2)^2}{2 B_e} \right) & k\ge k'
\end{array}
\right.\nn\\
&\equiv& e^{\ii\frac{(s p^2 + s' p'^2)l^1}{2 B_e}}
\mathcal{G}_{k k'}^{s s'}\left( l^1,s' p'^2 - s p^2 \right)
\label{eq:Jkk2}
\eea
\end{widetext}
In this last expression, we defined the function
\begin{widetext}
\bea
\mathcal{G}_{k k'}^{s s'}\left( l^1,s' p'^2 - s p^2 \right) = 2\pi e^{-\frac{(l^1)^2 + (s' p'2 - s p^2)^2}{4 B_e}}\left\{
\begin{array}{cc}
\sqrt{\frac{k!}{k'!}}\left(\ii\frac{  l^1 +\ii (s' p'^2 - s p^2) }{\sqrt{2 B_e}}\right)^{k' - k} L_{k}^{k'-k}\left( \frac{(l^1)^2 + (s' p'^2 - s p^2)^2}{2 B_e} \right) & k'\ge k\\
\sqrt{\frac{k'!}{k!}}\left(\ii\frac{  l^1 - \ii(s' p'^2 - s p^2) }{\sqrt{2 B_e}}\right)^{k - k'} L_{k'}^{k-k'}\left( \frac{(l^1)^2 + (s' p'^2 - s p^2)^2}{2 B_e} \right) & k\ge k'
\end{array}
\right.
\eea
\end{widetext}
This last expression can be further simplified, by applying the identity of the Laguerre polynomials
\bea
\frac{(-z)^m}{m!} L_n^{m-n}(z) = \frac{(-z)^n}{n!}L_m^{n-m}(z),
\eea
such that
\begin{widetext}
\bea
&&\sqrt{\frac{k'!}{k!}}\left(\ii\frac{  l^1 -\ii (s' p'^2 - s p^2) }{\sqrt{2 B_e}}\right)^{k - k'} L_{k'}^{k-k'}\left( \frac{(l^1)^2 + (s' p'^2 - s p^2)^2}{2 B_e} \right) = \ii^{k'-k}\sqrt{\frac{k!}{k'!}}\left( \frac{(l^1)^2 + (s' p'^2 - s p^2)^2}{2 B_e}\right)^{\frac{k'-k}{2}}\nn\\
&&\times\left( \frac{l^1 +\ii (s' p'^2 - s p^2)  }{\sqrt{(l^1)^2 + (s' p'^2 - s p^2)^2} } \right)^{k-k'} L_k^{k'-k}\left( \frac{(l^1)^2 + (s' p'^2 - s p^2)^2}{2 B_e} \right)
\eea
\end{widetext}
Finally, defining the angle
\bea
\tan\phi_{\perp} = \frac{s' p'^2 - s p^2}{l^1},
\eea
and correspondingly
\bea
\frac{l^1 \pm\ii (s' p'^2 - s p^2)  }{\sqrt{(l^1)^2 + (s' p'^2 - s p^2)^2} } &=& \cos(\phi_{\perp}) \pm \ii\sin(\phi_{\perp})\nn\\
&=& e^{\pm\ii\phi_{\perp}},
\eea
we finally get
\bea
\mathcal{G}_{k k'}^{s s'}\left( l^1,s' p'^2 - s p^2 \right) &=& 2\pi e^{-\frac{(l^1)^2 + (s' p'2 - s p^2)^2}{4 B_e}} \sqrt{\frac{k!}{k'!}}\ii^{k'-k}e^{\ii(k' - k)\phi_{\perp}}\left( \frac{(l^1)^2 + (s' p'^2 - s p^2)^2}{2 B_e} \right)^{\frac{k'-k}{2}}\nonumber\\
&\times&L_k^{k'-k}\left( \frac{(l^1)^2 + (s' p'^2 - s p^2)^2}{2 B_e} \right)\nn
\eea

We further notice that this expression can be interpreted by defining the \lq\lq perp" 2-component vector
\bea
l_{\perp} = (l^1,s'p'^2 - s p^2)\equiv (l^1,l^2)
\eea
such that the function above becomes
\bea
\mathcal{G}_{k k'}^{s s'}\left( l^1,l^2 = s' p'^2 - s p^2 \right) = 2\pi e^{-\frac{l_{\perp}^2}{4 B_e}} \sqrt{\frac{k!}{k'!}}\ii^{k'-k}e^{\ii(k' - k)\phi_{\perp}}\left( \frac{l_{\perp}^2}{2 B_e} \right)^{\frac{k'-k}{2}}  L_k^{k'-k}\left( \frac{l_{\perp}^2}{2 B_e} \right)\nn
\eea


\section{The next to leading order contribution to the vertex}
\label{NLO}
In the appendix we demonstrate Eq.~\eqref{eq_F2Full}. We start by the
definition given in Eq.~\eqref{general}
\bea
(2\pi)^3 \delta^{(3)}( \tilde{l} - \tilde{p} + \tilde{p}' )\mathcal{T}^{\mu}_{\lambda\lambda'}(\overline{p},\overline{p}',l) = \int_{x,y,z} \, \gamma^{\nu} S_F(x,y) \gamma^{\mu} S_F(y,z) \gamma^{\alpha} D_{\alpha\nu}(z,x) e^{-\ii l\cdot y} \mathcal{F}_e^*(x,\overline{p}_{\lambda}) \mathcal{F}_e(z,\overline{p}'_{\lambda'}).
\eea
In order to compute this last expression, we shall apply the Ritus eigenbasis representation for the Fermion propagators defined in Eq.~\eqref{eq_Ritusprop}, such that
\bea
S_F(x,y) &=& \sum_{n}\int \frac{d^3 \tilde{q}}{(2\pi)^3} \mathbb{E}^e(x,\tilde{q},n)\frac{\slashed{\Pi}_s(q_{\parallel},n) + m}{q_{\parallel}^2 - 2 n B_e - m^2}\tilde{\mathbb{E}}^{e}(y,\tilde{q},n)\nn\\
&=& \sum_{n}\int \frac{d^3 \tilde{q}}{(2\pi)^3} \sum_{\omega,\omega'}\Delta^{\omega}\frac{\slashed{\Pi}_s(q_{\parallel},n) + m}{q_{\parallel}^2 - 2 n B_e - m^2}\Delta^{\omega'}\mathcal{F}_e(x,\overline{q}_{\omega})\mathcal{F}^{*}_e(y,\overline{q}_{\omega'})\nn\\
S_F(y,z) &=& \sum_{n'}\int \frac{d^3 \tilde{q}'}{(2\pi)^3} \sum_{\beta,\beta'}\Delta^{\beta}\frac{\slashed{\Pi}_s(q'_{\parallel},n') + m}{q_{\parallel}^{'2} - 2 n' B_e - m^2}\Delta^{\beta'}\mathcal{F}_e(y,\overline{q}'_{\beta})\mathcal{F}^{*}_e(z,\overline{q}'_{\beta'}),
\label{eq_Ritusprop}
\eea
and the corresponding expression for the photon propagator in the Feynman gauge
\bea
D_{\alpha\nu}(z,x) = \int \frac{d^4 r}{(2\pi)^4}\frac{(-\ii)g_{\alpha\nu}}{r^2 + \ii\epsilon}e^{-\ii (z - x)\cdot r}
\eea
Then, inserting the expressions above we obtain
\bea
(2\pi)^3 \delta^{(3)}( \tilde{l} - \tilde{p} + \tilde{p}' )\mathcal{T}^{\mu}_{\lambda\lambda'}(\overline{p},\overline{p}',l) &=& \sum_{n,n'} \int \frac{d^3 \tilde{q}}{(2\pi)^3} \int \frac{d^3 \tilde{q}'}{(2\pi)^3}\int \frac{d^4 r}{(2\pi)^4}\frac{(-\ii)}{r^2 + \ii\epsilon}\sum_{\omega,\omega',\beta,\beta'}\mathcal{I}_x\mathcal{I}_y\mathcal{I}_z\nn\\
&&\times\gamma^{\nu}\Delta^{\omega}\frac{\slashed{\Pi}_s(q_{\parallel},n) + m}{q_{\parallel}^2 - 2 n B_e - m^2}\Delta^{\omega'}\gamma^{\mu}\Delta^{\beta}\frac{\slashed{\Pi}_s(q'_{\parallel},n') + m}{q_{\parallel}^{'2} - 2 n' B_e - m^2}\Delta^{\beta'}\gamma_{\nu}.
\eea
In this last expression, we defined the integrals (see Appendix~\ref{IntegralsDD} for the analytical expressions)
\bea
\mathcal{I}_x &=& \int d^4 x e^{\ii r\cdot x} \mathcal{F}_e^{*}(x,\overline{p}_{\lambda})\mathcal{F}_e(x,\overline{q}_{\omega})= (2\pi)^3 \delta^{(3)}(-\tilde{r}- \tilde{p} + \tilde{q})\mathcal{J}_{k_{\lambda},n_{\omega}}^{--}(p^2,q^2,-r^1)\nn\\
\mathcal{I}_y &=& \int d^4 y e^{-\ii l\cdot y} \mathcal{F}_e^{*}(y,\overline{q}_{\omega'})\mathcal{F}_e(y,\overline{q}'_{\beta}) = (2\pi)^3\delta^{(3)}(\tilde{l}-\tilde{q}+\tilde{q}')\mathcal{J}_{n_{\omega'},n'_{\beta}}^{--}(q^2,q^{'2},l^1)
\nn\\
\mathcal{I}_z &=& \int d^4 z e^{-\ii r\cdot z} \mathcal{F}_e^{*}(z,\overline{q}'_{\beta'})\mathcal{F}_e(z,\overline{p}'_{\lambda'}) = (2\pi)^3\delta^{(3)}(\tilde{r}-\tilde{q}'+\tilde{p}')\mathcal{J}_{n'_{\beta'},k'_{\lambda'}}^{--}(q^{'2},p^{'2},r^1)
\eea
By means of the delta functions, we can integrate the momenta subject to the following constraints
\bea
\tilde{q} &=& \tilde{r} + \tilde{p}\nn\\
\tilde{q}' &=& \tilde{r} + \tilde{p}'\nn\\
\tilde{l} &=& \tilde{q} - \tilde{q}' = \tilde{p} - \tilde{p}'.
\eea
Thus, after cancelling the global factor $(2\pi)^3 \delta^{(3)}( \tilde{l} - \tilde{p} + \tilde{p}' )$ at both sides, we obtain
\bea
&&\mathcal{T}^{\mu}_{\lambda\lambda'}(\overline{p},\overline{p}',l) = 
\sum_{n,n'}\sum_{\omega,\omega',\beta,\beta'} \int \frac{d^4 r}{(2\pi)^4}\mathcal{J}_{k_{\lambda},n_{\omega}}^{--}(p^2,r^2 + p^2,-r^1)
\mathcal{J}_{n_{\omega'},n'_{\beta}}^{--}(r^2 + p^2,r^2 + p^{'2},l^1)\nn\\
&&\times\mathcal{J}_{n'_{\beta'},k'_{\lambda'}}^{--}(r^2 + p^{'2},p^{'2},r^1)
\frac{(-\ii)}{r^2 + \ii\epsilon}\gamma^{\nu}\Delta^{\omega}\frac{\slashed{\Pi}_s((p + r)_{\parallel},n) + m}{(p + r)_{\parallel}^2 - 2 n B_e - m^2}\Delta^{\omega'}\gamma^{\mu}\Delta^{\beta}\frac{\slashed{\Pi}_s((p' + r)_{\parallel},n') + m}{(p' + r)_{\parallel}^{2} - 2 n' B_e - m^2}\Delta^{\beta'}\gamma_{\nu}.
\eea
Now, we shall recombine the denominators by using Feynman parameters
\bea
&&\frac{1}{\left( r^2 + \ii\epsilon \right)\left( (p + r)_{\parallel}^2 - 2 n B_e - m^2 \right)\left( (p'+r)_{\parallel}^{'2} - 2 n' B_e - m^2 \right)}
\equiv 2\int_{x,y,z}\frac{\delta(x + y + z - 1)}{\left[h_{\parallel}^2 - D(r_{\perp}^2)\right]^3}
\eea
where we defined, after completing the square (and using the condition $x + y + z = 1$)
\bea
h_{\parallel} = r_{\parallel} + y p_{\parallel} + z p'_{\parallel},
\eea
and the function
\bea
D(r_{\perp}^2) = (y p_{\parallel} + z p'_{\parallel})^2 + 2( y n + z n' ) B_e + (y + z) m^2 - y p_{\parallel}^2 - z p_{\parallel}^{'2} + x r_{\perp}^2.
\label{eq_D2}
\eea
Substituting these relations, we have
\bea
&&\mathcal{T}^{\mu}_{\lambda\lambda'}(\overline{p},\overline{p}',l) = 
\sum_{n,n'}\sum_{\omega,\omega',\beta,\beta'}2\int_{x,y,z}\delta(x + y + z - 1)\nn\\
&&\times\int \frac{d^2 r_{\perp}}{(2\pi)^2}\mathcal{J}_{k_{\lambda},n_{\omega}}^{--}(p^2,r^2 + p^2,-r^1)
\mathcal{J}_{n_{\omega'},n'_{\beta}}^{--}(r^2 + p^2,r^2 + p^{'2},l^1)\mathcal{J}_{n'_{\beta'},k'_{\lambda'}}^{--}(r^2 + p^{'2},p^{'2},r^1)\nn\\
&&\times(-\ii)\int\frac{d^2 h_{\parallel}}{(2\pi)^2}
\frac{ \gamma^{\nu}\Delta^{\omega}\left[\slashed{\Pi}_s(h_{\parallel} +(1 - y) p_{\parallel} - z p'_{\parallel},n) + m\right]\Delta^{\omega'}\gamma^{\mu}\Delta^{\beta}\left[\slashed{\Pi}_s(h_{\parallel}  - y p_{\parallel} +(1 - z) p'_{\parallel},n') + m\right]\Delta^{\beta'}\gamma_{\nu}}{\left[h^2_{\parallel} - D(r_{\perp}^2)\right]^3}.
\eea
In what follows, we shall take care of the reduced expression for the numerator
\bea
Num &=& \gamma^{\nu}\Delta^{\omega}\left[\slashed{\Pi}_s(h_{\parallel} +(1 - y) p_{\parallel} - z p'_{\parallel},n) + m\right]\Delta^{\omega'}\gamma^{\mu}\Delta^{\beta}\left[\slashed{\Pi}_s(h_{\parallel}  - y p_{\parallel} +(1 - z) p'_{\parallel},n') + m\right]\Delta^{\beta'}\gamma_{\nu}\nn\\
&=& \gamma^{\nu}\Delta^{\omega}\left[ \slashed{h}_{\parallel} + \slashed{\Pi}((1-y)p_{\parallel}-zp'_{\parallel},n)+ m)\right]\Delta^{\omega'}\gamma^{\mu}\Delta^{\beta}\left[ \slashed{h}_{\parallel} + \slashed{\Pi}(-yp_{\parallel} + (1 - z)p'_{\parallel},n') + m\right]\Delta^{\beta'}\gamma_{\nu}
\eea
Expanding the product, we apply the integral identities
\bea
\int \frac{d^2 h_{\parallel}}{(2\pi)^2}\frac{1}{\left[ h_{\parallel}^2 - D(r_{\perp}^2)\right]^3}&=&  -\frac{\ii}{8\pi}D^{-2}(r_{\perp}^2)
\nn\\
\int \frac{d^2 h_{\parallel}}{(2\pi)^2}\frac{h_{\parallel}^{\alpha}h_{\parallel}^{\alpha'}}{\left[ h_{\parallel}^2 - D(r_{\perp}^2)\right]^3} 
&=&  \frac{\ii}{16\pi}g_{\parallel}^{\alpha\alpha'}D^{-1}(r_{\perp}^2),
\eea
and we further eliminate the linear terms in $h_{\parallel}$ that vanish upon integration. After this procedure, we have that the numerator reduces to the sum of the following terms
\bea
\int \frac{d^2 h_{\parallel}}{(2\pi)^2}\frac{Num}{\left[ h_{\parallel}^2 - D(r_{\perp}^2) \right]^3} = N_1 + N_2 + N_3 + N_4
\eea
Here, we define the following contributions
\bea
N_1 &=& g_{\parallel\alpha\alpha'}\gamma^{\nu}\Delta^{\omega}\gamma^{\alpha}_{\parallel}\Delta^{\omega'}\gamma^{\mu}\Delta^{\beta}\gamma^{\alpha'}_{\parallel}\Delta^{\beta'}\gamma_{\nu}\left( \frac{\ii}{16\pi}D^{-1}(r_{\perp}^2) \right)\nn\\
N_2 &=& \gamma^{\nu}\Delta^{\omega}\slashed{\Pi}((1-y)p_{\parallel}-zp'_{\parallel},n)\Delta^{\omega'}\gamma^{\mu}\Delta^{\beta}\slashed{\Pi}(-yp_{\parallel} + (1 - z)p'_{\parallel},n')\Delta^{\beta'}\gamma_{\nu}\left(  -\frac{\ii}{8\pi}D^{-2}(r_{\perp}^2) \right)\nn\\
N_3 &=& m\left[ \gamma^{\nu}\Delta^{\omega}\slashed{\Pi}((1-y)p_{\parallel}-zp'_{\parallel},n)\Delta^{\omega'}\gamma^{\mu}\Delta^{\beta}\Delta^{\beta'}\gamma_{\nu} + \gamma^{\nu}\Delta^{\omega}\Delta^{\omega'}\gamma^{\mu}\Delta^{\beta}\slashed{\Pi}(-yp_{\parallel} + (1 - z)p'_{\parallel},n')\Delta^{\beta'}\gamma_{\nu}\right]\left( - \frac{\ii}{8\pi}D^{-2}(r_{\perp}^2) \right)\nn\\
N_4 &=& m^2 \gamma^{\nu}\Delta^{\omega}\Delta^{\omega'}\gamma^{\mu}\Delta^{\beta}\Delta^{\beta'}\gamma_{\nu}\left(  -\frac{\ii}{8\pi}D^{-2}(r_{\perp}^2) \right)
\eea
With these definitions, we have that
\bea
&&\mathcal{T}^{\mu}_{\lambda\lambda'}(\overline{p},\overline{p}',l) = 
\sum_{n,n'}\sum_{\omega,\omega',\beta,\beta'}2\int\delta(x + y + z - 1)\nn\\
&&\times\int \frac{d^2 r_{\perp}}{(2\pi)^2}\mathcal{J}_{k_{\lambda},n_{\omega}}^{--}(p^2,r^2 + p^2,-r^1)
\mathcal{J}_{n_{\omega'},n'_{\beta}}^{--}(r^2 + p^2,r^2 + p^{'2},l^1)\mathcal{J}_{n'_{\beta'},k'_{\lambda'}}^{--}(r^2 + p^{'2},p^{'2},r^1)(-\ii)\left( N_1 + N_2 + N_3 + N_4 \right)\nn\\
\label{eq_taumuapp}
\eea

Given that in this work we want to discuss the anomalous magnetic form factor in the purely transverse direction, which appears exclusively in the $N_2$ factor, we concentrate on working out this term explicitly.

\subsection{The term $N_2$}

The $N_2$ term is given by
\bea
N_2&=&\gamma^\nu \Delta^\omega[(1-y)\slashed{p}_\parallel-z{\slashed{p}'}_{\!\!\parallel}+\slashed{\Pi}_\perp(n)]\Delta^{\omega^{\prime}} \gamma^\mu \Delta^\beta [-y\slashed{p}_\parallel + (1-z){\slashed{p}'}_{\!\!\parallel}+\slashed{\Pi}_\perp(n')]\Delta^{\beta^\prime}\gamma_\nu\nonumber\\
&=& \gamma^\nu \Delta^\omega[\Delta^{\omega^\prime}\{(1-y)\slashed{p}_\parallel - z{\slashed{p}'}_{\!\!\parallel}\}+\Delta^{-\omega^\prime}\slashed{\Pi}_\perp(n)]\gamma^\mu\Delta^\beta[\Delta^{\beta^\prime}\{-y\slashed{p}_\parallel + (1-z){\slashed{p}'}_{\!\!\parallel}\}+\Delta^{-\beta^\prime}\slashed{\Pi}_\perp(n')]\gamma_\nu\nonumber\\
&=&\gamma^\nu \Delta^\omega[\delta_{\omega^\prime,\omega}\{(1-y)\slashed{p}_\parallel - z{\slashed{p}'}_{\!\!\parallel}\}+\delta_{\omega^\prime,-\omega}\slashed{\Pi}_\perp(n)]\left(\Delta^{-\beta}\gamma^\mu_\perp+\Delta^\beta\gamma^\mu_\parallel\right)\nonumber\\
&\times&[\delta_{\beta^\prime,\beta}\{-y\slashed{p}_\parallel + (1-z){\slashed{p}'}_{\!\!\parallel}\}+\delta_{\beta^\prime,-\beta}\slashed{\Pi}_\perp(n')]\gamma_\nu\nonumber\\
&\equiv& A+B
\label{N2}
\eea
where 
\bea
A&\!\!=\!\!&\gamma^\nu \Delta^\omega[\delta_{\omega^\prime,\omega}\delta_{\omega,-\beta}\{(1-y)\slashed{p}_\parallel - z{\slashed{p}'}_{\!\!\parallel}\}+\delta_{\omega^\prime,-\omega}\delta_{\omega,\beta}\slashed{\Pi}_\perp(n)]\gamma^\mu_\perp[\delta_{\beta^\prime,\beta}\{-y\slashed{p}_\parallel + (1-z){\slashed{p}'}_{\!\!\parallel}\}+\delta_{\beta^\prime,-\beta}\slashed{\Pi}_\perp(n')]\gamma_\nu
\label{N2A}
\eea
and
\bea
B&\!\!=\!\!&\gamma^\nu \Delta^\omega[\delta_{\omega^\prime,\omega}\delta_{\omega,-\beta}\{(1-y)\slashed{p}_\parallel - z{\slashed{p}'}_{\!\!\parallel}\}+\delta_{\omega^\prime,-\omega}\delta_{\omega,\beta}\slashed{\Pi}_\perp(n)]\gamma^\mu_\parallel[\delta_{\beta^\prime,\beta}\{-y\slashed{p}_\parallel + (1-z){\slashed{p}'}_{\!\!\parallel}\}+\delta_{\beta^\prime,-\beta}\slashed{\Pi}_\perp(n')]\gamma_\nu.
\label{N2B}
\eea
Placing the spin projector all the way to the left, we explicitly obtain
\bea
A&=&[ \Delta^\omega\gamma^\nu_\parallel+ \Delta^{-\omega}\gamma^\nu_\perp][\delta_{\omega^\prime,\omega}\delta_{\omega,-\beta}\{\cdot\}_{\parallel\alpha}\gamma^\alpha_\parallel+\delta_{\omega^\prime,-\omega}\delta_{\omega,\beta}\Pi_{\perp\alpha}(n)\gamma^\alpha_\perp]\gamma^\mu_\perp[\delta_{\beta^\prime,\beta}\{\cdot\cdot\}_{\parallel\alpha^\prime}\gamma^{\alpha^\prime}_\parallel+\delta_{\beta^\prime,-\beta}\Pi_{\perp\alpha^\prime}(n')\gamma^{\alpha^\prime}_\perp]\gamma_\nu\nonumber\\
&=&\Delta^\omega[\delta_{\omega^\prime,\omega}\delta_{\omega,-\beta}\{\cdot\}_{\parallel\alpha}\gamma^\nu_\parallel\gamma^\alpha_\parallel+\delta_{\omega^\prime,-\omega}\delta_{\omega,\beta}\Pi_{\perp\alpha}(n)\gamma^\nu_\parallel\gamma^\alpha_\perp]\gamma^\mu_\perp[\delta_{\beta^\prime,\beta}\{\cdot\cdot\}_{\parallel\alpha^\prime}\gamma^{\alpha^\prime}_\parallel\gamma_\nu+\delta_{\beta^\prime,-\beta}\Pi_{\perp\alpha^\prime}(n')\gamma^{\alpha^\prime}_\perp\gamma_\nu]\nonumber\\
&+&\Delta^{-\omega}[\delta_{\omega^\prime,\omega}\delta_{\omega,-\beta}\{\cdot\}_{\parallel\alpha}\gamma^\nu_\perp\gamma^\alpha_\parallel+\delta_{\omega^\prime,-\omega}\delta_{\omega,\beta}\Pi_{\perp\alpha}(n)\gamma^\nu_\perp\gamma^\alpha_\perp]\gamma^\mu_\perp[\delta_{\beta^\prime,\beta}\{\cdot\cdot\}_{\parallel\alpha^\prime}\gamma^{\alpha^\prime}_\parallel\gamma_\nu+\delta_{\beta^\prime,-\beta}\Pi_{\perp\alpha^\prime}(n')\gamma^{\alpha^\prime}_\perp\gamma_\nu]\nonumber\\
&=&\Delta^\omega\left[\delta_{\omega^\prime,\omega}\delta_{\omega,-\beta}\delta_{\beta^\prime,\beta}\{\cdot\}_{\parallel\alpha}\{\cdot\cdot\}_{\parallel\alpha^\prime}\gamma^\nu_\parallel\gamma^\alpha_\parallel\gamma^\mu_\perp\gamma^{\alpha^\prime}_\parallel\gamma_\nu
+
\delta_{\omega^\prime,\omega}\delta_{\omega,-\beta}\delta_{\beta^\prime,-\beta}\{\cdot\}_{\parallel\alpha}\Pi_{\perp\alpha^\prime}(n')\gamma^\nu_\parallel\gamma^\alpha_\parallel\gamma^\mu_\perp\gamma^{\alpha^\prime}_\perp\gamma_\nu\right.\nonumber\\
&+&\left.
\delta_{\omega^\prime,-\omega}\delta_{\omega,\beta}\delta_{\beta^\prime,\beta}\Pi_{\perp\alpha}(n)\{\cdot\cdot\}_{\parallel\alpha^\prime}\gamma^\nu_\parallel\gamma^\alpha_\perp\gamma^\mu_\perp\gamma^{\alpha^\prime}_\parallel\gamma_\nu
+
\delta_{\omega^\prime,-\omega}\delta_{\omega,\beta}\delta_{\beta^\prime,-\beta}\Pi_{\perp\alpha}(n)\Pi_{\perp\alpha^\prime}(n')\gamma^\nu_\parallel\gamma^\alpha_\perp\gamma^\mu_\perp\gamma^{\alpha^\prime}_\perp\gamma_\nu
\right]\nonumber\\
&+&\Delta^{-\omega}\left[\delta_{\omega^\prime,\omega}\delta_{\omega,-\beta}\delta_{\beta^\prime,\beta}\{\cdot\}_{\parallel\alpha}\{\cdot\cdot\}_{\parallel\alpha^\prime}\gamma^\nu_\perp\gamma^\alpha_\parallel\gamma^\mu_\perp\gamma^{\alpha^\prime}_\parallel\gamma_\nu
+
\delta_{\omega^\prime,\omega}\delta_{\omega,-\beta}\delta_{\beta^\prime,-\beta}\{\cdot\}_{\parallel\alpha}\Pi_{\perp\alpha^\prime}(n')\gamma^\nu_\perp\gamma^\alpha_\parallel\gamma^\mu_\perp\gamma^{\alpha^\prime}_\perp\gamma_\nu\right.\nonumber\\
&+&\left.
\delta_{\omega^\prime,-\omega}\delta_{\omega,\beta}\delta_{\beta^\prime,\beta}\Pi_{\perp\alpha}(n)\{\cdot\cdot\}_{\parallel\alpha^\prime}\gamma^\nu_\perp\gamma^\alpha_\perp\gamma^\mu_\perp\gamma^{\alpha^\prime}_\parallel\gamma_\nu
+
\delta_{\omega^\prime,-\omega}\delta_{\omega,\beta}\delta_{\beta^\prime,-\beta}\Pi_{\perp\alpha}(n)\Pi_{\perp\alpha^\prime}(n')\gamma^\nu_\perp\gamma^\alpha_\perp\gamma^\mu_\perp\gamma^{\alpha^\prime}_\perp\gamma_\nu
\right]
\label{N2Aexpl}
\eea
where
\bea
\{\cdot\}_{\parallel\alpha}&\equiv&\{(1-y)p_\parallel - z p^\prime_\parallel\}_\alpha\nonumber\\
\{\cdot\cdot\}_{\parallel\alpha}&\equiv&\{-y p_\parallel + (1-z)p^\prime_\parallel\}_\alpha.
\label{dots}
\eea
The gamma matrix contraction gives
\begin{eqnarray}
(i)\ \ \ \gamma^{\nu}_\parallel\gamma^{\alpha}_\parallel\gamma^{\mu}_\perp\gamma^{\alpha^\prime}_\parallel\gamma_{\nu}&=&2\gamma^{\mu}_\perp\gamma^{\alpha^\prime}_\parallel\gamma^{\alpha}_\parallel\nonumber\\
(ii)\ \ \ \gamma^{\nu}_\parallel\gamma^{\alpha}_\parallel\gamma^{\mu}_\perp\gamma^{\alpha^\prime}_\perp\gamma_{\nu}&=&0\nonumber\\
(iii)\ \ \  \gamma^{\nu}_\parallel\gamma^{\alpha}_\perp\gamma^{\mu}_\perp\gamma^{\alpha^\prime}_\parallel\gamma_{\nu}&=&0\nonumber\\
(iv)\ \ \ \gamma^{\nu}_\parallel\gamma^{\alpha}_\perp\gamma^{\mu}_\perp\gamma^{\alpha^\prime}_\perp\gamma_{\nu}&=&-2\gamma^{\alpha}_\perp\gamma^{\mu}_\perp\gamma^{\alpha^\prime}_\perp\nonumber\\
(v)\ \ \ \gamma^{\nu}_\perp\gamma^{\alpha}_\parallel\gamma^{\mu}_\perp\gamma^{\alpha^\prime}_\parallel\gamma_{\nu}&=&0\nonumber\\
(vi)\ \ \  \gamma^{\nu}_\perp\gamma^{\alpha}_\parallel\gamma^{\mu}_\perp\gamma^{\alpha^\prime}_\perp\gamma_{\nu}&=&-2\gamma^{\alpha}_\parallel\gamma^{\alpha^\prime}_\perp\gamma^{\mu}_\perp\nonumber\\
(vii)\ \ \  \gamma^{\nu}_\perp\gamma^{\alpha}_\perp\gamma^{\mu}_\perp\gamma^{\alpha^\prime}_\parallel\gamma_{\nu}&=&-2\gamma^{\alpha^\prime}_\parallel\gamma^{\mu}_\perp\gamma^{\alpha}_\perp\nonumber\\
(viii)\ \ \ \gamma^{\nu}_\perp\gamma^{\alpha}_\perp\gamma^{\mu}_\perp\gamma^{\alpha^\prime}_\perp\gamma_{\nu}&=&0.
\label{contractionsN2}
\end{eqnarray}
Therefore
\bea
A&=&-2\left\{\Delta^\omega\left(\delta_{\omega^\prime,\omega}\delta_{\omega,-\beta}\delta_{\beta^\prime,\beta}[-y\slashed{p}_\parallel + (1-z)\slashed{p}^\prime_\parallel]\gamma^\mu_\perp[(1-y)\slashed{p}_\parallel - z\slashed{p}^\prime_\parallel]+
\delta_{\omega^\prime,-\omega}\delta_{\omega,\beta}\delta_{\beta^\prime,-\beta}\slashed{\Pi}_\perp(n)\gamma^\mu_\perp\slashed{\Pi}_\perp(n^\prime)\right)
\right.\nonumber\\
&+&\left.
\Delta^{-\omega}\left(
\delta_{\omega^\prime,\omega}\delta_{\omega,-\beta}\delta_{\beta^\prime,-\beta}\slashed{\Pi}_\perp(n^\prime)\gamma^\mu_\perp[(1-y)\slashed{p}_\parallel - z\slashed{p}^\prime_\parallel]+\delta_{\omega^\prime,-\omega}\delta_{\omega,\beta}\delta_{\beta^\prime,\beta}[-y\slashed{p}_\parallel +(1-z)\slashed{p}^\prime_\parallel]\gamma^\mu_\perp\slashed{\Pi}_\perp(n)
\right)
\right\}.
\eea
Similarly
\bea
\!\!B&\!\!\!=\!\!\!&
[\Delta^\omega\gamma^\nu_\parallel + \Delta^{-\omega}\gamma^\nu_\perp][\delta_{\omega^\prime,\omega}\delta_{\omega,\beta}\{\cdot\}_{\parallel\alpha}\gamma^\alpha_\parallel + \delta_{\omega,-\beta}\delta_{\omega^\prime,-\omega}\Pi_{\perp\alpha}(n)\gamma^\alpha_\perp]\gamma^\mu_\parallel [\delta_{\beta^\prime,\beta}\{\cdot\cdot\}_{\parallel\alpha^\prime}\gamma^{\alpha^\prime}_\parallel\gamma_\nu + \delta_{\beta^\prime,-\beta}\Pi_{\perp\alpha^\prime}(n^\prime)\gamma^{\alpha^\prime}_\perp\gamma_\nu]\nonumber\\
&=&\Delta^\omega\left\{
\delta_{\omega^\prime,\omega}\delta_{\omega,\beta}\delta_{\beta^\prime,\beta}\{\cdot\}_{\parallel\alpha}\{\cdot\cdot\}_{\parallel\alpha^\prime}\gamma^\nu_\parallel\gamma^\alpha_\parallel\gamma^\mu_\parallel\gamma^{\alpha^\prime}_\parallel\gamma_\nu + 
\delta_{\omega^\prime,\omega}\delta_{\omega,\beta}\delta_{\beta^\prime,-\beta}\{\cdot\}_{\parallel\alpha}\Pi_{\perp\alpha^\prime}(n^\prime)\gamma^\nu_\parallel\gamma^\alpha_\parallel\gamma^\mu_\parallel\gamma^{\alpha^\prime}_\perp\gamma_\nu\right.\nonumber\\
&+&\left.
\delta_{\omega,-\beta}\delta_{\omega^\prime,-\omega}\delta_{\beta^\prime,\beta}\Pi_{\perp\alpha}(n)\{\cdot\cdot\}_{\parallel\alpha^\prime}\gamma^\nu_\parallel\gamma^\alpha_\perp\gamma^\mu_\parallel\gamma^{\alpha^\prime}_\parallel\gamma_\nu
+
\delta_{\omega,-\beta}\delta_{\omega^\prime,-\omega}\delta_{\beta^\prime,-\beta}\Pi_{\perp\alpha}(n)\Pi_{\perp\alpha^\prime}(n^\prime)\gamma^\nu_\parallel\gamma^\alpha_\perp\gamma^\mu_\parallel\gamma^{\alpha^\prime}_\perp\gamma_\nu
\right\}\nonumber\\
&+&
\Delta^{-\omega}\left\{
\delta_{\omega^\prime,\omega}\delta_{\omega,\beta}\delta_{\beta^\prime,\beta}\{\cdot\}_{\parallel\alpha}\{\cdot\cdot\}_{\parallel\alpha^\prime}\gamma^\nu_\perp\gamma^\alpha_\parallel\gamma^\mu_\parallel\gamma^{\alpha^\prime}_\parallel\gamma_\nu
+
\delta_{\omega^\prime,\omega}\delta_{\omega,\beta}\delta_{\beta^\prime,-\beta}\{\cdot\}_{\parallel\alpha}\Pi_{\perp\alpha^\prime}(n^\prime)\gamma^\nu_\perp\gamma^\alpha_\parallel\gamma^\mu_\parallel\gamma^{\alpha^\prime}_\perp\gamma_\nu\right.\nonumber\\
&+&\left.
\delta_{\omega,-\beta}\delta_{\omega^\prime,-\omega}\delta_{\beta^\prime,\beta}\Pi_{\perp\alpha}(n)\{\cdot\cdot\}_{\parallel\alpha^\prime}\gamma^\nu_\perp\gamma^\alpha_\perp\gamma^\mu_\parallel\gamma^{\alpha^\prime}_\parallel\gamma_\nu
+
\delta_{\omega,-\beta}\delta_{\omega^\prime,-\omega}\delta_{\beta^\prime,-\beta}\Pi_{\perp\alpha}(n)\Pi_{\perp\alpha^\prime}(n^\prime)\gamma^\nu_\perp\gamma^\alpha_\perp\gamma^\mu_\parallel\gamma^{\alpha^\prime}_\perp\gamma_\nu
\right\},
\eea
and the corresponding gamma matrix contractions are
\begin{eqnarray}
(i)\ \ \ \gamma^{\nu}_\parallel\gamma^{\alpha}_\parallel\gamma^{\mu}_\parallel\gamma^{\alpha^\prime}_\parallel\gamma_{\nu}&=&0\nonumber\\
(ii)\ \ \ \gamma^{\nu}_\parallel\gamma^{\alpha}_\parallel\gamma^{\mu}_\parallel\gamma^{\alpha^\prime}_\perp\gamma_{\nu}&=&-2\gamma^{\alpha^\prime}_\perp\gamma^\mu_\parallel\gamma^\alpha_\parallel\nonumber\\
(iii)\ \ \  \gamma^{\nu}_\parallel\gamma^{\alpha}_\perp\gamma^{\mu}_\parallel\gamma^{\alpha^\prime}_\parallel\gamma_{\nu}&=&-2\gamma^{\alpha^\prime}_\parallel\gamma^\mu_\parallel\gamma^\alpha_\perp\nonumber\\
(iv)\ \ \ \gamma^{\nu}_\parallel\gamma^{\alpha}_\perp\gamma^{\mu}_\parallel\gamma^{\alpha^\prime}_\perp\gamma_{\nu}&=&0\nonumber\\
(v)\ \ \ \gamma^{\nu}_\perp\gamma^{\alpha}_\parallel\gamma^{\mu}_\parallel\gamma^{\alpha^\prime}_\parallel\gamma_{\nu}&=&-2\gamma^{\alpha}_\parallel\gamma^\mu_\parallel\gamma^{\alpha^\prime}_\parallel\nonumber\\
(vi)\ \ \  \gamma^{\nu}_\perp\gamma^{\alpha}_\parallel\gamma^{\mu}_\parallel\gamma^{\alpha^\prime}_\perp\gamma_{\nu}&=&0\nonumber\\
(vii)\ \ \  \gamma^{\nu}_\perp\gamma^{\alpha}_\perp\gamma^{\mu}_\parallel\gamma^{\alpha^\prime}_\parallel\gamma_{\nu}&=&0\nonumber\\
(viii)\ \ \ \gamma^{\nu}_\perp\gamma^{\alpha}_\perp\gamma^{\mu}_\parallel\gamma^{\alpha^\prime}_\perp\gamma_{\nu}&=&-2\gamma^{\alpha^\prime}_\perp\gamma^\mu_\parallel\gamma^\alpha_\perp.
\label{contractionsN2B}
\end{eqnarray}
Therefore
\bea
B&=&-2\left\{\Delta^\omega\left(
\delta_{\omega^\prime,\omega}
\delta_{\omega,\beta}
\delta_{\beta^\prime,-\beta}
\slashed{\Pi}_\perp(n^\prime)
\gamma^\mu_\parallel
[(1-y)\slashed{p}_\parallel - z\slashed{p}^\prime_\parallel]
+
\delta_{\omega,-\beta}
\delta_{\omega^\prime,-\omega}
\delta_{\beta^\prime,\beta}
[-y\slashed{p}_\parallel +(1-z)\slashed{p}^\prime_\parallel]
\gamma^\mu_\parallel
\slashed{\Pi}_\perp(n)\right)
\right.\nonumber\\
&+&\left.
\Delta^{-\omega}\left(
\delta_{\omega^\prime,\omega}
\delta_{\omega,\beta}
\delta_{\beta^\prime,\beta}
[(1-y)\slashed{p}_\parallel 
-z\slashed{p}^\prime_\parallel]
\gamma^\mu_\parallel
[-y\slashed{p}_\parallel +(1-z)\slashed{p}^\prime_\parallel]
+
\delta_{\omega,-\beta}
\delta_{\omega^\prime,-\omega}
\delta_{\beta^\prime,-\beta}
\slashed{\Pi}_\perp(n^\prime)
\gamma^\mu_\parallel
\slashed{\Pi}_\perp(n)
\right)
\right\}.
\eea

Now, we need to compute the sandwich of $A$ and $B$ between the external spin projectors and the Dirac spinors, in other words, we need to compute
 \bea
 \overline{u}_e(p_\parallel,k,a)\Delta^\lambda N_2 \Delta^{\lambda^\prime}u_e(p_\parallel^\prime,k^\prime,a^\prime).
 \label{tocompute}
 \eea
 Applying the generalized form of the Gordon identity discussed in Appendix~\ref{Gordon}, the result is
 \bea
 \overline{u}_e(p_\parallel,k,a)\Delta^\lambda N_2 \Delta^{\lambda^\prime}u_e(p_\parallel^\prime,k^\prime,a')&=&-2\overline{u}(p_\parallel,k)\Delta^\lambda\Big\{
 \delta_{\lambda^\prime,-\lambda}\delta_{\omega,\lambda}\delta_{\omega^\prime,\omega}\delta_{\omega,-\beta}\delta_{\beta,\beta^\prime}\Big[\Big\{
 \left(m^2(y(1-y)-z(1-z)
 \right)\nonumber\\
 &+&2B_e\left(ky(1-y)-k^\prime z(1-z)\right)-2(1-z)(1-y)p_\parallel\cdot p_\parallel^\prime\Big\}\gamma_\perp^\mu\nonumber\\
 &-& (1-y-z)\Big(m^2\gamma^\mu_\perp-m\slashed{\Pi}_\perp(k)\gamma^\mu_\perp
 -m\gamma^\mu_\perp\slashed{\Pi}_\perp(k^\prime)+\slashed{\Pi}_\perp(k)\gamma^\mu_\perp\slashed{\Pi}_\perp(k^\prime)\Big)\Big]\nonumber\\
 &+&\delta_{\omega,-\lambda}\delta_{\lambda^\prime,\lambda}\delta_{\omega^\prime,\omega}\delta_{\omega,-\beta}\delta_{\beta^\prime,-\beta}\Big[m(1-y-z)\slashed{\Pi}_\perp(n^\prime)\gamma^\mu_\perp - (1-y)\slashed{\Pi}(n^\prime)\gamma^\mu_\perp\slashed{\Pi}_\perp(k)\nonumber\\
 &+& z\slashed{\Pi}(n^\prime)\gamma^\mu_\perp\slashed{\Pi}_\perp(k^\prime)\Big]\nonumber\\
 &+& \delta_{\lambda^\prime,-\lambda}\delta_{\omega,\lambda}\delta_{\omega^\prime,-\omega}\delta_{\omega,\beta}\delta_{\beta^\prime,-\beta}\slashed{\Pi}_\perp(n)\gamma^\mu_\perp\slashed{\Pi}_\perp(n^\prime)\nonumber\\
 &+&\delta_{\omega,-\lambda}\delta_{\lambda^\prime,\lambda}\delta_{\omega^\prime,-\omega}\delta_{\omega,\beta}\delta_{\beta^\prime,\beta}\Big[m(1-y-z)\gamma^\mu_\perp\slashed{\Pi}_\perp(n) + y\slashed{\Pi}_\perp(k)\gamma^\mu_\perp\slashed{\Pi}_\perp(n)\nonumber\\
 &-& (1-z)\slashed{\Pi}_\perp(k^\prime)\gamma^\mu_\perp\slashed{\Pi}_\perp(n)\Big]\nonumber\\
 &+&\delta_{\omega,-\beta}\delta_{\omega^\prime,-\omega}\delta_{\beta^\prime,\beta}\delta_{\omega,\lambda}\delta_{\lambda,-\lambda^\prime}\Big[\Big( m(z-y-1)\gamma^\mu_\parallel + y\slashed{\Pi}_\perp(k)\gamma^\mu_\parallel\Big)\slashed{\Pi}_\perp(n)\nonumber\\
 &+& (1-z)\gamma^\mu_\parallel\slashed{\Pi}_\perp(k^\prime)\slashed{\Pi}_\perp(n) + 2(1-z)p^{\prime\mu}_\parallel\slashed{\Pi}_\perp(n)\Big]\nonumber\\
 &+&\delta_{\omega,\lambda}\delta_{\lambda,-\lambda^\prime}\delta_{\omega^\prime,\omega}\delta_{\omega,\beta}\delta_{\beta^\prime,-\beta}\Big[-m(1-y+z)\slashed{\Pi}_\perp(n^\prime)\gamma^\mu_\parallel + (1-y)\slashed{\Pi}_\perp(n^\prime)\slashed{\Pi}_\perp(k)\gamma^\mu_\parallel\nonumber\\
 &-&z\slashed{\Pi}_\perp(n^\prime)\slashed{\Pi}_\perp(k^\prime)\gamma^\mu_\parallel\Big]\nonumber\\
 &+&\delta_{\omega,-\lambda}\delta_{\lambda,\lambda^\prime}\delta_{\omega^\prime,\omega}\delta_{\omega,\beta}\delta_{\beta^\prime,\beta}\Big[(1-y-z)\left(m-\slashed{\Pi}_\perp(k)\right)\gamma^\mu_\parallel\left(m-\slashed{\Pi}_\perp(k^\prime\right)\nonumber\\
 &+& z(1-z)\gamma^\mu_\parallel\left(m^2 + 2k^\prime B_e\right) - y(1-y)\left(m^2 + 2kB_e\right)\gamma^\mu_\parallel\nonumber\\
 &+& 2(yz p^{\prime\mu}_\parallel - y(1-y)p^\mu_\parallel)\left(m-\slashed{\Pi}_\perp(k)\right)\nonumber\\
 &+& 2\left(yzp^\mu_\parallel - z(1-z)p^{\prime\mu}_\parallel\right)\left(m-\slashed{\Pi}_\perp(k^\prime)\right) - 2yz p^{\prime}_\parallel\cdot p_\parallel\gamma^\mu_\parallel\Big]\nonumber\\
 &+&\delta_{\omega,-\lambda}\delta_{\lambda,\lambda^\prime}\delta_{\omega,-\beta}\delta_{\omega^\prime,-\omega}\delta_{\beta^\prime,-\beta}\Big[\slashed{\Pi}_\perp(n^\prime)\gamma^\mu_\parallel\slashed{\Pi}_\perp(n)\Big]\Big\}u(p^\prime_\parallel,k^\prime)
 \label{sandwichN2}
 \eea
 Now we extract the form factors as coefficients of the vector structures. 
 \bea
 \overline{u}_e(p_\parallel,k,a)\Delta^\lambda N_2 \Delta^{\lambda^\prime}u_e(p_\parallel^\prime,k^\prime,a^\prime)
 &=&-2\overline{u}_e(p_\parallel,k,a)\Delta^\lambda\Big\{
 \mathcal{F}^{\lambda,\lambda'}_{2\perp\perp}\sigma^{\mu 2}_{\perp\perp}
 +\ldots\Big\}u_e(p^\prime_\parallel,k^\prime,a^\prime),
 \label{FFN2}
 \eea
 where the $\ldots$ represent structures that we are not yet computing but that do not contribute to the purely transverse anomalous magnetic moment and will be  reported on elsewhere soon. 
 We find that this contribution becomes
 \bea
 \mathcal{F}_{2\perp\perp}^{\lambda\lambda'}&=&im\sqrt{2B_e}(1-y-z)\Big\{
 \delta_{\lambda^\prime,-\lambda}\delta_{\omega,\lambda}\delta_{\omega^\prime,\omega}\delta_{\omega,-\beta}\delta_{\beta,\beta^\prime}(\sqrt{k_\lambda} - \sqrt{k^\prime_{\lambda'}})\nonumber\\
 &+&\delta_{\omega,-\lambda}\delta_{\lambda^\prime,\lambda}\delta_{\omega^\prime,\omega}\delta_{\omega,-\beta}\delta_{\beta^\prime,-\beta}\sqrt{n^\prime}\nonumber\\
 &-&\delta_{\omega,-\lambda}\delta_{\lambda^\prime,\lambda}\delta_{\omega^\prime,-\omega}\delta_{\omega,\beta}\delta_{\beta^\prime,\beta}\sqrt{n}\Big\}.
 \label{FFN2expl}
 \eea
Inserting this expression into the full phase-space integration involved in Eq.~\eqref{eq_taumuapp} for the NLO vertex, we have that the corresponding form factor is given by
\begin{eqnarray}
F_{2\perp\perp}(p,p',l) &=& \frac{ 2m (-i)}{8\pi} \sqrt{2 B_e}
\sum_{n,n'}\sum_{\omega,\omega'}\sum_{\beta,\beta'}\Bigg\{ 
\delta_{\lambda',-\lambda}\delta_{\omega,\lambda}\delta_{\omega',\omega}\delta_{\omega,-\beta}\delta_{\beta',\beta}(\sqrt{k_{\lambda}} - \sqrt{k'_{\lambda'}})\nonumber\\ 
&&+ \delta_{\omega,-\lambda}\delta_{\lambda',\lambda}\delta_{\omega',\omega}\delta_{\omega,-\beta}\delta_{\beta',-\beta}\sqrt{n'}- \delta_{\omega,-\lambda}\delta_{\lambda',\lambda}\delta_{\omega',-\omega}\delta_{\omega,\beta}\sqrt{n}
\Bigg\}\int dx dy dz\, \delta(x + y + z - 1) \cdot x\nonumber\\
&&\cdot \int\frac{d^2 r_{\perp}}{(2\pi)^2}\mathcal{J}^{--}_{k_{\lambda},n_{\omega}}\left( p^2,r^2 + p^2,-r^1 \right)
\mathcal{J}_{n_{\omega'},n'_{\beta}}^{--}\left( r^2 + p^2,r^2 + p^{'2},l^1 \right)\mathcal{J}_{n'_{\beta'},k'_{\lambda'}}^{--}\left( r^2 + p^{'2},p^{'2},r^1 \right) D^{-2}(r_{\perp}^2)\nonumber\\
\end{eqnarray}

The angular integral involving the product of three functions is explicitly calculated in Appendix~\ref{Angular_integral_NLO}.


\section{A generalized form of the Gordon identity}
\label{Gordon}
As encountered in several expressions along the manuscript, we have linear combinations of vector components of the canonical momenta, in structures of the type $\overline{u}_e(p_{\parallel},n,a)\Delta^{\omega}(\Pi^{\mu} + \Pi^{'\mu})u_e(p'_\parallel,n',a')$, that need to be substituted in favor of other explicit tensor components in order to identify contributions to the form factors. For this purpose, we here prove a generalization of the Gordon identity as follows
\bea
&&\overline{u}_e(p_{\parallel},n,a) \left[ \Delta^{\omega}\left( \Pi^{\mu}_{\parallel} + \Pi^{'\mu}_{\parallel} \right) + \Delta^{\omega}\Pi^{'\mu}_{\perp} + \Delta^{-\omega}\Pi^{\mu}_{\perp}\right]u_e(p'_{\parallel},n',a')\nn\\
&&=\overline{u}_e(p_{\parallel},n,a)
\left[
\Delta^{\omega}\left( 2 m \gamma^{\mu} + \ii \sigma^{\mu\alpha}_{\parallel\parallel}\left( \Pi'_{\alpha} - \Pi_{\alpha} \right) + \ii \sigma^{\mu\alpha}_{\perp\parallel} \left( \Pi'_{\alpha} - \Pi_{\alpha} \right) + \ii\sigma^{\mu\alpha}_{\parallel\perp}\Pi'_{\alpha} + \ii\sigma_{\perp\perp}^{\mu\alpha}\Pi'_{\alpha}\right)\right.\nn\\
&&\left.- \Delta^{-\omega}\left( \ii\sigma_{\parallel\perp}^{\mu\alpha}\Pi_{\alpha} + \ii\sigma_{\perp\perp}^{\mu\alpha}\Pi_{\alpha} \right)
\right]
u_e(p'_{\parallel},n',a').
\label{eq_Gordon}
\eea
For the proof, we start by considering the Ritus version of the Dirac equation for the spinor representing the fermion and its conjugate
\bea
\left[ \slashed{\Pi}(p'_{\parallel},n') - m \right]u_e(p'_{\parallel},n',a') &=& 0\nn\\
\overline{u}_e(p_{\parallel},n,a)\left[
\slashed{\Pi}(p_{\parallel},n) - m\right] &=& 0
\eea
Let us now multiply the first equation above, from the left, by $\overline{u}_e(p_{\parallel},n,a) \Delta^{\omega}\gamma^{\mu}\rightarrow$, to obtain (using the shorthand notation $\slashed{\Pi}(p'_{\parallel},n') \equiv \slashed{\Pi}'$)
\bea
\overline{u}_e(p_{\parallel},n,a)\left[ \Delta^{\omega}\gamma^{\mu}\slashed{\Pi}' - m\Delta^{\omega}\gamma^{\mu}\right] u_e(p'_{\parallel},n',a') = 0,
\label{eq_G1}
\eea
and similarly, multiplying the second equation from the right by $\leftarrow \Delta^{\omega}\gamma^{\mu}u_e (p_{\parallel},n,a)$, we obtain
\bea
\overline{u}_e(p_{\parallel},n,a)\left[ \slashed{\Pi}\Delta^{\omega}\gamma^{\mu} - m\Delta^{\omega}\gamma^{\mu}\right] u_e(p'_{\parallel},n',a') = 0.
\label{eq_G2}
\eea
By adding up Eq.~\eqref{eq_G1} and Eq.~\eqref{eq_G2}, we end up with
\bea
\overline{u}_e(p_{\parallel},n,a)\left[ \Delta^{\omega}\gamma^{\mu}\slashed{\Pi}'+\slashed{\Pi}\Delta^{\omega}\gamma^{\mu} - 2 m\Delta^{\omega}\gamma^{\mu}\right] u_e(p'_{\parallel},n',a') = 0.
\eea
Finally, we expand the Lorentz product in the expression above
\bea
\overline{u}_e(p_{\parallel},n,a)\left[ \Delta^{\omega}\gamma^{\mu}\gamma^{\alpha}\Pi'_{\alpha}+\Pi_{\alpha}\gamma^{\alpha}\Delta^{\omega}\gamma^{\mu} - 2 m\Delta^{\omega}\gamma^{\mu}\right] u_e(p'_{\parallel},n',a') = 0.
\eea
We further use the property
\bea
\gamma^{\alpha}\Delta^{\omega} = \Delta^{\omega}\gamma^{\alpha}_{\parallel} + \Delta^{-\omega}\gamma^{\alpha}_{\perp},
\eea
and organize the terms according to the two different spin projections
\bea
\overline{u}_e(p_{\parallel},n,a)\left[ \Delta^{\omega}\left(\gamma^{\mu}\gamma^{\alpha}\Pi'_{\alpha} + \gamma^{\alpha}_{\parallel}\gamma^{\mu}\Pi_{\alpha}\right)+\Pi_{\alpha}\Delta^{-\omega}\gamma^{\alpha}_{\perp}\gamma^{\mu} - 2 m\Delta^{\omega}\gamma^{\mu}\right] u_e(p'_{\parallel},n',a') = 0.
\label{eq_G3}
\eea
By using the identity,
\bea
\gamma^{\mu}\gamma^{\alpha} = \frac{1}{2}\left\{\gamma^{\mu},\gamma^{\alpha} \right\} + \frac{1}{2}\left[\gamma^{\mu},\gamma^{\alpha} \right] = g^{\mu\alpha} - \ii \sigma^{\mu\alpha} = g^{\mu\alpha}_{\parallel} + g^{\mu\alpha}_{\perp} - \ii \sigma^{\mu\alpha}_{\parallel\parallel} - \ii \sigma^{\mu\alpha}_{\parallel\perp} - \ii \sigma^{\mu\alpha}_{\perp\perp} - \ii \sigma^{\mu\alpha}_{\perp\parallel}
\eea
and similarly
\bea
\gamma^{\alpha}_{\parallel}\gamma^{\mu} = \gamma^{\alpha}_{\parallel}\gamma^{\mu}_{\parallel} + \gamma^{\alpha}_{\parallel}\gamma^{\mu}_{\perp} = g^{\mu\alpha}_{\parallel} + \ii\sigma_{\parallel\parallel}^{\mu\alpha}+\ii\sigma_{\perp\parallel}^{\mu\alpha},
\eea
we insert those into Eq.~\eqref{eq_G3}, and reorganizing the terms we arrive at the final expression Eq.~\eqref{eq_Gordon}.
\section{Integral over Feynman parameters}
\label{Int_Feyn}
In this section, we show the details on how to compute the integral over Feynman parameters, Eq.~\eqref{eq_Ieta}, and defined by
\bea
\mathcal{K}(\eta) = \int \int \int dx\,dy\,dz \frac{x \delta(x + y + z - 1)}{\left[ \left( y p_{\parallel} + z p'_{\parallel} \right)^2 + 2\left( y n + z n'  \right) B_e + (y + z)m^2 - y p_{\parallel}^2 - z p_{\parallel}^{'2} + 2 B_e \eta x \right]^2}
\eea
As a first step, we remark that the factor of $x$ in the numerator may be generated by a parametric differentiation with respect to $\eta$, as follows
\bea
\mathcal{K}(\eta) &=& - (2 B_e)^{-1}\frac{\partial}{\partial\eta}
\int \int \int dx\,dy\,dz \frac{ \delta(x + y + z - 1)}{\left[ \left( y p_{\parallel} + z p'_{\parallel} \right)^2 + 2\left( y n + z n'  \right) B_e + (y + z)m^2 - y p_{\parallel}^2 - z p_{\parallel}^{'2} + 2 B_e \eta x \right]}\nn\\
&=& - (2 B_e)^{-1}\frac{\partial}{\partial\eta}\int_0^1 dy\int_0^{1-y} dz \frac{ 1}{\left[ \left( y p_{\parallel} + z p'_{\parallel} \right)^2 + 2\left( y n + z n'  \right) B_e + (y + z)m^2 - y p_{\parallel}^2 - z p_{\parallel}^{'2} + 2 B_e \eta x \right]},
\eea
where in the second step, we carried over the explicit integration over $x$.
Now, we factor out the denominator as a polynomial in $z$, as follows
\bea
\Delta(p'_{\parallel},p_{\parallel})&=&\left( y p_{\parallel} + z p'_{\parallel} \right)^2 + 2\left( y n + z n'  \right) B_e + (y + z)m^2 - y p_{\parallel}^2 - z p_{\parallel}^{'2} + 2 B_e \eta (1 - y - z))\nonumber\\
&=& p^{'2}_{\parallel}\left( z^2 + B z + C \right) = p^{'2}_{\parallel} \left( z - z_1 \right)\left( z - z_2 \right),
\eea
where we defined
\bea
B &=& \frac{2 y p'_{\parallel}\cdot p_{\parallel} + 2 n' B_e + m^2 - p^{'2}_{\parallel} - 2 B_e \eta }{p_{\parallel}^{'2}}\nn\\
C &=& \frac{ y^2 p_{\parallel}^2 + y(2 (n - \eta) B_e + m^2 - p_{\parallel}^2) + 2 B_e  \eta}{p_{\parallel}^{'2} }
\eea
where we defined the roots
\bea
z_1 &=& - \frac{B}{2} +\frac{1}{2}\sqrt{B^2 - 4 C} \nn\\ 
z_2 &=& - \frac{B}{2} - \frac{1}{2} \sqrt{B^2 - 4 C}.
\eea
In the particular case when we take the limit $l_{\parallel} = p'_{\parallel} - p_{\parallel} \rightarrow 0$, the two roots reduce to a simpler expression
\bea
&&z_1 \rightarrow a + b y + \sqrt{y \frac{(2 n' - 2 n ) B_e}{p_{\parallel}^2} - c}\nonumber\\
&&z_2 \rightarrow a + b y - \sqrt{y \frac{(2 n' - 2 n ) B_e}{p_{\parallel}^2} - c}, 
\eea
where we defined the momentum-dependent parameters
\bea
a &=& -\frac{2 (n' - \eta)B_e  + m^2 - p_{\parallel}^2}{2 p_{\parallel}^2}\nn\\
b &=& -1 \nn\\
c &=& \frac{2 B_e }{p_{\parallel}^2} \eta - \frac{\left( 2 (n' - \eta) B_e + m^2 - p_{\parallel}^2 \right)^2}{4 \left(p_{\parallel}^2\right)^2}.
\label{eq_Apabc}
\eea
We need to distinguish two different cases: Case (1) $n' = n$, and case (2) $n' \ne n$.
\subsection{Case (1): $n' = n$}
Here, the roots further reduce to the simplified forms:
\bea
&&z_1 \rightarrow a + \sqrt{-c} + b y\nn\\
&&z_2 \rightarrow a - \sqrt{-c} + b y
\eea
Substituting, we thus have
\bea
\mathcal{K}(\eta) &=& - (2 B_e p_{\parallel}^{2})^{-1}\frac{\partial}{\partial\eta}\int_0^1 dy\int_0^{1-y} \frac{dz}{(z - z_1)(z - z_2)}\nn\\
&=& -\frac{1}{2 B_e p_{\parallel}^{2}}\frac{\partial}{\partial\eta}\int_0^1 \frac{dy}{z_1 - z_2}\ln\left[ \frac{(1 - y - z_1) z_2}{z_1(1 - y - z_2)} \right]\nn\\
&=& -\frac{1}{2 B_e p_{\parallel}^{2}}\frac{\partial}{\partial\eta}\int_0^1 \frac{dy}{2\sqrt{ - c}}
\left[ \ln(z_2) + \ln(1 - y - z_1) - \ln z_1 - \ln (1 - y - z_2) \right]\nn\\
&=& -\frac{1}{2 B_e p_{\parallel}^{2}}\frac{\partial}{\partial\eta}\int_0^1 \frac{dy}{2\sqrt{ - c}}
\Bigg[ \ln(a + b y -\sqrt{-c}) + \ln(1 - a - (b + 1)y - \sqrt{ - c} ) \nn\\ 
&&-\ln( a + b y + \sqrt{- c} ) - \ln(1 - a - (1 + b)y + \sqrt{ - c})
\Bigg]\nonumber\\
&=& -\frac{1}{2 B_e p_{\parallel}^{2}}\frac{\partial}{\partial\eta}\int_0^1 \frac{dy}{2\sqrt{ - c}}
\Bigg[ \ln(a - y -\sqrt{-c}) + \ln(1 - a - \sqrt{ - c} ) \nn\\ 
&&-\ln( a - y + \sqrt{- c} ) - \ln(1 - a  + \sqrt{ - c})
\Bigg],\nonumber\\
\eea
where we used that $b=-1$. By further using the simple identity
\bea
\int_0^1 dy \ln(A - y) = A \ln (A)- (A-1)\ln (A-1) - 1,
\eea
we have for this case
\bea
\mathcal{K}(\eta) &=& -\frac{1}{2 B_e p_{\parallel}^{2}}\frac{\partial}{\partial\eta}\left[ \frac{1}{2\sqrt{-c}}\left( \ln(1 - a - \sqrt{ - c} )  - \ln(1 - a  + \sqrt{ - c})
+ (1 - a - \sqrt{ - c} ) \ln (1 - a - \sqrt{ - c} )\right.\right.\nn\\
&&\left.\left.- ( - a - \sqrt{ - c} )\ln ( - a - \sqrt{ - c} ) - (1 - a + \sqrt{ - c} ) \ln (1 - a + \sqrt{ - c} )
+ ( - a + \sqrt{ - c} )\ln ( - a + \sqrt{ - c} )
\right) \right].
\eea
\subsection{Case (2) $n' \ne n$}
Let us define the parameter
\bea
W = \frac{(2 n' - 2 n ) B_e}{p_{\parallel}^2} 
\label{eq_ApW}
\eea
Substituting, we thus have
\bea
\mathcal{K}(\eta) &=& - (2 B_e p_{\parallel}^{2})^{-1}\frac{\partial}{\partial\eta}\int_0^1 dy\int_0^{1-y} \frac{dz}{(z - z_1)(z - z_2)}\nn\\
&=& -\frac{1}{2 B_e p_{\parallel}^{2}}\frac{\partial}{\partial\eta}\int_0^1 \frac{dy}{z_1 - z_2}\ln\left[ \frac{(1 - y - z_1) z_2}{z_1(1 - y - z_2)} \right]\nn\\
&=& -\frac{1}{2 B_e p_{\parallel}^{2}}\frac{\partial}{\partial\eta}\int_0^1 \frac{dy}{2\sqrt{W y - c}}
\left[ \ln(z_2) + \ln(1 - y - z_1) - \ln z_1 - \ln (1 - y - z_2) \right]\nn\\
&=& -\frac{1}{2 B_e p_{\parallel}^{2}}\frac{\partial}{\partial\eta}\int_0^1 \frac{dy}{2\sqrt{W y - c}}
\Bigg[ \ln(a + b y -\sqrt{W y-c}) + \ln(1 - a - (b + 1)y - \sqrt{W y - c} ) \nn\\ 
&&-\ln( a + b y + \sqrt{W y - c} ) - \ln(1 - a - (1 + b)y + \sqrt{W y - c})
\Bigg].
\eea
Each of the integrals in the expression above are of the form (for suitable choices of parameters $A,B,C$)
\bea
I^{\pm}[A,B,C,W] = \int_{0}^{1}\frac{dy}{2\sqrt{W y - C}}\ln(A + B y \pm \sqrt{W y - C}),
\label{eq_GINT}
\eea
such that
\bea
\mathcal{K}(\eta) &=& 
-\frac{1}{2 B_e p_{\parallel}^{2}}\frac{\partial}{\partial\eta}\left\{ I^{-}[a,b,c,W] + I^{-}[1-a,-(1+b),c,W] - I^{+}[a,b,c,W]  
- I^{+}[1-a,-(1+b),c,W]
\right\}\nn\\
&=& -\frac{1}{2 B_e p_{\parallel}^{2}}\frac{\partial}{\partial\eta}\left\{ I^{-}[a,-1,c,W] + I^{-}[1-a,0,c,W] - I^{+}[a,-1,c,W]  
- I^{+}[1-a,0,c,W]
\right\},
\eea
where in the second line we used $b = -1$, and the parameters $a,c$ defined in Eq.~\eqref{eq_Apabc}, with $W$ in Eq.~\eqref{eq_ApW}.
Therefore, we proceed at last to evaluate the generic integral Eq.~\eqref{eq_GINT}. 

We need to consider two different cases: $B \ne 0$ and $B = 0$, respectively. 

Let us first consider $B \ne 0$. We introduce the change of variables $y \rightarrow W y$, such that
\bea
I^{\pm}[A,B,C,W] = \frac{1}{W}\int_{0}^{W}\frac{dy}{2\sqrt{ y - C}}\ln\left(A + \frac{B}{W} y \pm \sqrt{ y - C}\right).
\eea
Let us further introduce the change of variables
\bea
u &=& \sqrt{ y - C} \Rightarrow y = u^2 + C\nn\\
du &=& \frac{dy}{2\sqrt{y - C}},
\eea
such that
\bea
I^{\pm}[A,B,C,W] &\!\!=\!\!& \frac{1}{W}\int_{\sqrt{-C}}^{\sqrt{W-C}} du\,\ln\left( A + \frac{B C}{W} + \frac{B}{W} u^2 \pm u \right)\nn\\
&\!\!=\!\!& \frac{1}{W}\int_{\sqrt{-C}}^{\sqrt{W-C}} du\,\ln\left( \frac{B}{W} (u - u_1)(u - u_2) \right)\nn\\
&\!\!=\!\!& \frac{1}{W}\int_{\sqrt{-C}}^{\sqrt{W-C}} du \left[ \ln (B/W) + \ln(u - u_1) + \ln(u - u_2) \right]\nn\\
&\!\!=\!\!& \frac{1}{W}\Bigg[\left( \sqrt{W - C} - \sqrt{-C} \right)\ln (B/W) + \left( \sqrt{W - C} - u_1 \right)\ln(\sqrt{W - C} - u_1) - \left( \sqrt{W - C} - u_1 \right)\nn\\
&&- (\sqrt{-C} - u_1)\ln(\sqrt{-C} - u_1) + (\sqrt{-C} - u_1) + \left( \sqrt{W - C} - u_2 \right)\ln(\sqrt{W - C} - u_2)\nn\\ 
&&- \left( \sqrt{W - C} - u_2 \right)- (\sqrt{-C} - u_1)\ln(\sqrt{-C} - u_2) + (\sqrt{-C} - u_2)\Bigg],
\eea
where we define the roots of the quadratic equation $A + B C/W + (B/W) u^2 \pm u = 0$ as
\bea
u_1 &=& \mp \frac{W}{B} + \frac{W}{B}\sqrt{1 - 4 \frac{B}{W} (A + (B C)/W)}\nn\\
u_2 &=& \mp \frac{W}{B} - \frac{1}{B}\sqrt{1 - 4 \frac{B}{W} (A + (B C)/W)}.
\eea
Let us now consider the case $B = 0$, such that
\bea
\!\!\!\!I^{\pm}[A,0,C,W] &\!\!=\!\!& \frac{1}{W}\int_{0}^{W}\frac{dy}{2\sqrt{ y - C}}\ln(A \pm \sqrt{ y - C})\nn\\
&\!\!=\!\!& \frac{1}{W}\int_{\sqrt{-C}}^{\sqrt{W-C}} du\,\ln\left( A  \pm u \right)\nn\\
&\!\!=\!\!& \left.\frac{1}{W}\left[ (u \pm A)\ln(u \pm A) - u  \right]\right|_{\sqrt{-C}}^{\sqrt{W-C}}\nn\\
&\!\!=\!\!& \frac{1}{W}\left[ (\sqrt{W-C}\pm A)\ln(\sqrt{W-C}\pm A) - \sqrt{W-C} - (\sqrt{-C}\pm A)\ln(\sqrt{-C}\pm A) + \sqrt{-C}  \right].
\eea
\section{The angular integral involved in the vertex at one-loop (NLO)}
\label{Angular_integral_NLO}
We now present the details to obtain the angular integral (for $b=1,2$)
\bea
&&\int\frac{d^2 r_{\perp}}{(2\pi)^2}\mathcal{J}^{--}_{k_{\lambda},n_{\omega}}\left( p^2,r^2 + p^2,-r^1 \right)
\mathcal{J}_{n_{\omega'},n'_{\beta}}^{--}\left( r^2 + p^2,r^2 + p^{'2},l^1 \right)\mathcal{J}^{--}_{n'_{\beta'},k'_{\lambda'}}\left( r^2 + p^{'2},p^{'2},r^1 \right) D^{-b}(r_{\perp}^2)\nn\\ 
&&= (2\pi)^2 \sqrt{\frac{k_{\lambda}!n_{\omega'}!n'_{\beta'}!}{n_{\omega}!n'_{\beta}!k'_{\lambda'}!}}
 e^{\ii\left( n_{\omega} - k_{\lambda} + n'_{\beta} - n_{\omega'} + k'_{\lambda'} - n'_{\beta'}  \right)\left(\varphi_{l} + \pi\right)}e^{-\frac{l_{\perp}^2}{4 B_e}}\left( \frac{l_{\perp}^2}{2 B_e} \right)^{\frac{n'_{\beta} - n_{\omega'}}{2}}\nonumber\\
 &&\times L_{n_{\omega'}}^{n'_{\beta} - n_{\omega'}}\left( \frac{l_{\perp}^2}{2 B_e} \right)e^{-\ii\frac{\left( p^2 + p^{'2} \right) l^1}{2 B_e}}\mathcal{S}_{n_{\omega}+k'_{\lambda'}-k_{\lambda}-n'_{\beta'}}\nn\\
&&\times\int_{0}^{\infty}dr_{\perp} r_{\perp} e^{-\frac{r_{\perp}^2}{2 B_e}}\left( \frac{r_{\perp}^2}{2 B_e} \right)^{\frac{n_{\omega} + k'_{\lambda'} - k_{\lambda} - n'_{\beta'}}{2}}
L_{k_{\lambda}}^{n_{\omega} - k_{\lambda}}\left( \frac{r_{\perp}^2}{2 B_e} \right) L_{n'_{\beta'}}^{k'_{\lambda'} - n'_{\beta'}}\left( \frac{r_{\perp}^2}{2 B_e} \right)\nonumber\\
&&\times J_{\left| n_{\omega} + k'_{\lambda'} - k_{\lambda} - n'_{\beta'} \right|}\left( \sqrt{\frac{2}{B_e}}l_{\perp}r_{\perp} \right) D^{-b}(r_{\perp}^2)\nn\\
&&=(2\pi)^2 \sqrt{\frac{k_{\lambda}!n_{\omega'}!n'_{\beta'}!}{n_{\omega}!n'_{\beta}!k'_{\lambda'}!}} e^{\ii\left( n_{\omega} - k_{\lambda} + n'_{\beta} - n_{\omega'} + k'_{\lambda'} - n'_{\beta'}  \right)\left(\varphi_{l} + \pi\right)} e^{-\frac{l_{\perp}^2}{4 B_e}}  \left( \frac{l_{\perp}^2}{2 B_e} \right)^{\frac{n'_{\beta} - n_{\omega'}}{2}}\nonumber\\
&&\times L_{n_{\omega'}}^{n'_{\beta} - n_{\omega'}}\left( \frac{l_{\perp}^2}{2 B_e} \right)e^{-\ii\frac{\left( p^2 + p^{'2} \right) l^1}{2 B_e}}\mathcal{S}_{n_{\omega}+k'_{\lambda'}-k_{\lambda}-n'_{\beta'}}\nn\\
&&\times B_e\int_{0}^{\infty}d\eta\,e^{-\eta}\,\eta^{\frac{n_{\omega} + k'_{\lambda'} - k_{\lambda} - n'_{\beta'}}{2}}
L_{k_{\lambda}}^{n_{\omega} - k_{\lambda}}\left( \eta\right) L_{n'_{\beta'}}^{k'_{\lambda'} - n'_{\beta'}}\left( \eta \right) J_{\left| n_{\omega} + k'_{\lambda'} - k_{\lambda} - n'_{\beta'}\right|}\left( 2 l_{\perp}\sqrt{\eta} \right) D^{-b}\left( 2 B_e \eta \right),
\label{eq_AI3fact}
\eea
where we have made the change of variable $\eta=r_\perp^2/2B_e$ and 
\bea
D(2 B_e \eta) = (y p_{\parallel} + z p'_{\parallel})^2 + 2( y n + z n' ) B_e + (y + z) m^2 - y p_{\parallel}^2 - z p_{\parallel}^{'2} + 2 B_e x\, \eta.
\label{eq_DD2}
\eea
Here, we defined the signature
\bea
\mathcal{S}_{n_{\omega}+k'_{\lambda'}-k_{\lambda}-n'_{\beta'}} = \delta_{\left| n_{\omega}+k'_{\lambda'}-k_{\lambda}-n'_{\beta'} \right|,2 j} - \sgn\left( n_{\omega}+k'_{\lambda'}-k_{\lambda}-n'_{\beta'} \right)\delta_{\left| n_{\omega}+k'_{\lambda'}-k_{\lambda}-n'_{\beta'} \right|,2 j  + 1}
\eea
For the proof, we start writing the explicit expression for each of the tree factors in the integrand
\bea
\mathcal{J}^{--}_{k_{\lambda},n_{\omega}}(p^2,r^2+p^2,-r^1)
&=& e^{-\ii\frac{\left( r^2 + 2 p^2 \right)}{2 B_e}(-r^1)} \mathcal{G}_{k_{\lambda},n_{\omega}}^{--}(-r^1,-r^2)\nn\\
&=& (2\pi)\sqrt{\frac{k_{\lambda}!}{n_{\omega}!}}\ii^{n_{\omega} - k_{\lambda}} e^{-\frac{r_{\perp}^2}{4 B_e}} \left( \frac{r_{\perp}^2}{2 B_e} \right)^{\frac{n_{\omega} - k_{\lambda}}{2}}L_{k_{\lambda}}^{n_{\omega} - k_{\lambda}}\left( \frac{r_{\perp}^2}{2 B_e} \right) e^{\ii\left( n_{\omega} - k_{\lambda} \right)\phi_{\perp}} e^{\ii\frac{\left( r^2 + 2 p^2 \right)}{2 B_e}r^1}, 
\eea
with $\tan(\phi_{\perp}) = r^2/r^1$.
Similarly,
\bea
\mathcal{J}^{--}_{n_{\omega'},n'_{\beta}}(r^2 + p^2,r^2+p^{'2},l^1)
&=& e^{-\ii\frac{\left( 2 r^2 +  p^2 + p^{'2} \right)}{2 B_e}l^1} \mathcal{G}_{n_{\omega'},n'_{\beta}}^{--}(l^1,p^2 - p^{'2} = l^2)\nn\\
&=& (2\pi)\sqrt{\frac{n_{\omega'}!}{n'_{\beta}!}}\ii^{n'_{\beta} - n_{\omega'}} e^{-\frac{l_{\perp}^2}{4 B_e}}  \left( \frac{l_{\perp}^2}{2 B_e} \right)^{\frac{n'_{\beta} - n_{\omega'}}{2}}L_{n_{\omega'}}^{n'_{\beta} - n_{\omega'}}\left( \frac{l_{\perp}^2}{2 B_e} \right) e^{\ii\left( n'_{\beta} - n_{\omega'} \right)\varphi_{l}} e^{-\ii\frac{\left( 2 r^2 + p^2 + p^{'2} \right)}{2 B_e}l^1}\nn\\
\eea
where we defined the polar angle in the direction of the incident photon by $\tan\varphi_l = l^2/l^1$.
Finally, the third factor reduces to
\bea
\mathcal{J}^{--}_{n'_{\beta'},k'_{\lambda'}}(r^2 + p^{'2},p^{'2},r^1)
&=& e^{-\ii\frac{\left( r^2 + 2 p^{'2} \right)}{2 B_e}r^1} \mathcal{G}_{n'_{\beta'},k'_{\lambda'}}^{--}(r^1,r^2)\nn\\
&=& (2\pi)\sqrt{\frac{n'_{\beta'}!}{k'_{\lambda'}!}}\ii^{k'_{\lambda'} - n'_{\beta'}} e^{-\frac{r_{\perp}^2}{4 B_e}}  \left( \frac{r_{\perp}^2}{2 B_e} \right)^{\frac{k'_{\lambda'} - n'_{\beta'}}{2}}L_{n'_{\beta'}}^{k'_{\lambda'} - n'_{\beta'}}\left( \frac{r_{\perp}^2}{2 B_e} \right) e^{\ii\left( k'_{\lambda'} - n'_{\beta'} \right)\phi_{\perp}} e^{-\ii\frac{\left( r^2 + 2 p^{'2} \right)}{2 B_e}r^1}\nn\\
\eea
When assembling the product of all three factors, let us first simplify the phase
\bea
e^{\ii\left[\left( r^2 + 2 p^2 \right)r^1 - \left( 2 r^2 + p^2 + p'2 \right)l^1 - \left( r^2 + 2 p^{'2} \right)r^1 \right]/(2 B_e)} &=& e^{-\ii\frac{\left( p^2 + p'^2 \right)l^1}{2 B_e}} e^{\ii\left[ 2(p^2 - p^{'2})r^1 - 2 r^1 l^1\right]/(2 B_e)}\nn\\
&=& e^{-\ii\frac{\left( p^2 + p'^2 \right)l^1}{(2 B_e)}}
e^{2\ii\frac{\left( l^2 r^1 - l^1 r ^2 \right)}{(2 B_e)}},
\eea
where in the last step we applied the overall delta for momentum conservation $l^2 = p^2 - p^{'2}$. In order to proceed, let us now express the components of $(r^1 = r_{\perp}\cos\phi_{\perp},r^2 = r_{\perp}\sin\phi_{\perp})$, and $(l^1 = l_{\perp}\cos\varphi_l,l^2 = l_{\perp}\sin\varphi_l)$, such that
\bea
l^2 r^1 - l^1 r^2 = l_{\perp}r_{\perp}\left( \cos\phi_{\perp}\sin\varphi_l - \cos\varphi_l\sin\phi_{\perp} \right)= -l_{\perp}r_{\perp}\sin\left( \phi_{\perp} - \varphi_l \right)
\eea
Therefore, when choosing the integration coordinates it is convenient to define $\phi'_{\perp} = \phi_{\perp} - \varphi_l$, and correspondingly the integration measure $d^2 r_{\perp} = r_{\perp}dr_{\perp} d\phi'_{\perp}$, with $0\le\phi'_{\perp} < 2\pi$, and $0\le r_{\perp} < \infty$. Let us first integrate the angular coordinate, such that we need to evaluate
\bea
\int_0^{2\pi} d\phi_{\perp}' e^{\ii\left( n_{\omega} + k'_{\lambda'} - k_{\lambda} - n'_{\beta'} \right)\phi_{\perp}}e^{-\frac{\ii}{B_e}r_{\perp}l_{\perp}\sin\phi'_{\perp}}.
\eea
For this purpose, we use the well known Bessel-Fourier series expansion for a plane-wave
\bea
e^{-\ii z \sin\phi} = J_0(z) + 2\sum_{n=1}^{\infty}J_{2n}(z)\cos(2n\phi) - 2\ii\sum_{n=0}^{\infty}J_{2n+1}(z)\sin((2n+1)\phi).
\eea
Using the Euler formulae
\bea
2\cos(2n\phi) &=& e^{\ii 2n\phi} + e^{-\ii 2n\phi}\nn\\
2\ii\sin((2n+1)\phi) &=& e^{\ii 2n\phi} - e^{-\ii 2n\phi},
\eea
as well as the elementary identity
\bea
\int_0^{2\pi}d\phi e^{\ii (m - m')\phi} = 2\pi \delta_{m,m'},
\eea
we can integrate the series as follows
\bea
\int_0^{2\pi}d\phi e^{\ii m \phi - \ii z \sin\phi} &=& 2\pi\delta_{m,0}J_0(z) + 2\pi\sum_{n=1}^{\infty}\left(  \delta_{m,-2n} + \delta_{m,2n} \right)J_{2n}(z) - 2\pi\sum_{n=0}^{\infty}\left( \delta_{m,-(2n+1)} - \delta_{m,2n+1} \right)J_{2n+1}(z)\nn\\
&=& 2\pi J_{|m|}(z) \left( \delta_{|m|,2 j } - \left( \delta_{m,-(2j+1)} - \delta_{m,2j+1} \right)  \right)\nn\\
&=& 2\pi J_{|m|}(z) \left( \delta_{|m|,2 j } - \sgn(m)\delta_{|m|,2 j + 1 }\right)\nn\\
&\equiv& \mathcal{S}_m 2\pi J_{|m|}(z),
\eea
where here we used the dummy index $j\ge 0$ to indicate whether $|m| = 2 j$ is an even positive integer, or $|m| = 2j+1$ and odd positive integer, respectively, and the signature function
\bea
\mathcal{S}_m \equiv \left( \delta_{|m|,2 j } - \sgn(m)\delta_{|m|,2 j + 1 }\right).
\eea
Therefore, applying this result, we have the analytical expression
\bea
\int_0^{2\pi} d\phi_{\perp}' e^{\ii\left( n_{\omega} + k'_{\lambda'} - k_{\lambda} - n'_{\beta'} \right)\phi_{\perp}}e^{-\frac{\ii}{B_e}r_{\perp}l_{\perp}\sin\phi'_{\perp}}
= \mathcal{S}_{n_{\omega} + k'_{\lambda'} - k_{\lambda} - n'_{\beta'}}  2\pi J_{|n_{\omega} + k'_{\lambda'} - k_{\lambda} - n'_{\beta'}|}\left(  l_{\perp} r_{\perp}/B_e\right ). 
\eea
Inserting this result, and collecting the remaining factors, we arrive at the final expression stated in Eq.~\eqref{eq_AI3fact}.
\bibliographystyle{apsrev4-1}

\end{document}